\documentclass[12pt, authoryear]{elsarticle}

\makeatletter
\def\ps@pprintTitle{%
 \let\@oddhead\@empty
 \let\@evenhead\@empty
 \def\@oddfoot{\centerline{\thepage}}%
 \let\@evenfoot\@oddfoot}
\makeatother

\usepackage{amsmath}
\usepackage{graphicx}
\usepackage{verbatim}
\usepackage{subfigure}
\usepackage{color}
\usepackage{setspace}
\usepackage{float}
\usepackage{natbib}
\usepackage{footnote}
\makesavenoteenv{tabular}
\makesavenoteenv{table}

\usepackage[top=2.3cm,left=2.3cm,right=2.3cm,bottom=2.3cm]{geometry}

\begin{document}

\begin{frontmatter}

\title{On the surface composition of Triton's southern latitudes}

\author[lasp,irtf]{B. J. Holler}
\author[swri,irtf]{L. A. Young}
\author[low,irtf]{W. M. Grundy}
\author[swri,irtf]{C. B. Olkin}

\address[lasp]{Laboratory for Atmospheric and Space Physics, University of Colorado at Boulder, 1234 Innovation Dr., Boulder, CO 80303.}
\address[irtf]{Visiting or remote observer at the Infrared Telescope Facility, which is operated by the University of Hawaii under Cooperative Agreement \#NNX-08AE38A with the National Aeronautics and Space Administration, Science Mission Directorate, Planetary Astronomy Program.}
\address[swri]{Southwest Research Institute, 1050 Walnut St. \#300, Boulder, CO 80302.}
\address[low]{Lowell Observatory, 1400 W. Mars Hill Rd., Flagstaff, AZ 86001.}

\begin{abstract}
\onehalfspacing{We present the results of an investigation to determine the longitudinal (zonal) distributions and temporal evolution of ices on the surface of Triton. Between 2002 and 2014, we obtained 63 nights of near-infrared (0.67-2.55 $\mu$m) spectra using the SpeX instrument at NASA's Infrared Telescope Facility (IRTF). Triton has spectral features in this wavelength region from N$_2$, CO, CH$_4$, CO$_2$, and H$_2$O. Absorption features of ethane (C$_2$H$_6$) and $^{13}$CO are coincident at 2.405 $\mu$m, a feature that we detect in our spectra. We calculated the integrated band area (or fractional band depth in the case of H$_2$O) in each nightly average spectrum, constructed longitudinal distributions, and quantified temporal evolution for each of the chosen absorption bands. The volatile ices (N$_2$, CO, CH$_4$) show significant variability over one Triton rotation and have well-constrained longitudes of peak absorption. The non-volatile ices (CO$_2$, H$_2$O) show poorly-constrained peak longitudes and little variability. The longitudinal distribution of the 2.405 $\mu$m band shows little variability over one Triton rotation and is 97$\pm$44$^{\circ}$ and 92$\pm$44$^{\circ}$ out of phase with the 1.58 $\mu$m and 2.35 $\mu$m CO bands, respectively. This evidence indicates that the 2.405 $\mu$m band is due to absorption from non-volatile ethane. CH$_4$ absorption increased over the period of the observations while absorption from all other ices showed no statistically signifcant change. We conclude from these results that the southern latitudes of Triton are currently dominated by non-volatile ices and as the sub-solar latitude migrates northwards, a larger quantity of volatile ice is coming into view.}
\end{abstract}


\end{frontmatter}

\section{Introduction}
\onehalfspace{The images of Triton taken during the Voyager 2 flyby in August 1989 provided a tantalizing view of a diverse and dynamic surface. Two adjacent terrrains on the imaged hemisphere appear drastically different in both color and texture, hinting that these two distinct surface units may have different surface ice compositions (Stone and Miner, 1989). Triton is frequently compared to Pluto, and with good reason. Triton and Pluto are comparable in size and may have come from the same initial population; it is believed that Triton is a captured Kuiper Belt Object (Agnor and Hamilton, 2006). However, Triton has a more diverse collection of surface ices, with N$_2$, CO, CH$_4$, CO$_2$, and H$_2$O ices definitively identified in ground-based spectra (Cruikshank et al., 1993; Cruikshank et al., 2000). Spectral signatures, possibly due to C$_2$H$_6$ (ethane), are present as well (DeMeo et al., 2010). Two independent investigations suggest that Triton's surface temperature is about 38 K (Broadfoot et al., 1989; Tryka et al, 1993). At this temperature, the sublimation pressures of N$_2$ (22 $\mu$bar), CO (3 $\mu$bar), and CH$_4$ (0.002 $\mu$bar) are non-negligible, while the sublimation pressures of CO$_2$ (10$^{-24}$ $\mu$bar), H$_2$O (0 $\mu$bar), and ethane (2$\times$10$^{-16}$ $\mu$bar) are negligible (Fray and Schmitt, 2009). Henceforth, we will refer to N$_2$, CO, and CH$_4$ as volatile ices due to their relatively higher sublimation pressures compared to the non-volatile ices: CO$_2$, H$_2$O, and ethane. CO$_2$ and H$_2$O are less mobile and presumably constitute the substrate upon which the volatile ices (N$_2$, CO, CH$_4$) deposit. The presence of volatile ices on the surface, though predicted for an object with the surface temperature and diameter of Triton (Schaller and Brown, 2007; Johnson et al., 2015), is surprising. The capture process and subsequent circularization of Triton's orbit through tidal interactions heated Triton significantly, resulting in the production of a thick atmosphere and blowoff of volatile species (McKinnon et al., 1995). Present-day surface composition is puzzling in light of the dynamical history of Triton.\\
\indent The continued presence of volatile ices allows for an atmosphere around Triton. Despite its low surface temperature and high geometric albedo (0.719; Hicks and Buratti, 2004), the surface pressure on Triton was measured at 14$\pm$1 $\mu$bar by Voyager 2 in 1989 (Gurrola, 1995). Stellar occultations in the 1990s (through 1997) showed a surprising increase in both temperature and pressure (Elliot et al., 1998). The next occultation was observed in 2008 but the data have yet to be reduced and analyzed, so the current state of Triton's atmosphere is unknown (Sicardy et al., 2008). Triton's atmosphere is dominated by N$_2$ with traces of CO and CH$_4$ (Tyler et al., 1989; Lellouch et al., 2010). Volatile transport, driven by migration of the sub-solar point, is responsible for the observed spatial distributions and temporal evolution of surface ices on Triton (Buratti et al., 1994; Bauer et al., 2010). Previously published papers describe in more detail the significant seasonal variations on Triton due to migration of the sub-solar point (e.g., Trafton, 1984; Hansen and Paige, 1992; Moore and Spencer, 1990). Fig. 1 shows the change in the sub-solar latitude over the period 1000-3000 C.E. A subset of that figure covering 1980-2020 C.E. is presented in Fig. 2. After reaching its maximum southern extent in 2000 (-50$^{\circ}$), the sub-solar point\renewcommand{\thefootnote}{\fnsymbol{footnote}}\footnote{Triton is in a synchronous, retrograde orbit about Neptune. Its rotation is also retrograde with sunrise in the west. The sub-Neptune point is at 0$^{\circ}$ longitude. The south pole is the current summer pole.} has turned northward, reaching -42$^{\circ}$ at the time of this writing in mid-2015. Images obtained by Voyager 2 were taken during this extended period of southern illumination. A higher albedo in the southern hemisphere led some to argue for a south polar region covered in volatile ices (e.g., Stone and Miner, 1989; Moore and Spencer, 1990). With limited spectral information in the near-infrared from Voyager 2, the composition of various regions of Triton's surface from the flyby epoch are unknown.\\
\indent As measured by Voyager 2, the mixing ratio of CH$_4$ to N$_2$ in Triton's atmosphere is on the order of $\sim$10$^{-4}$ (Broadfoot et al., 1989; Herbert and Sandel, 1991). The presence of CH$_4$, even at these levels, drives ongoing photochemistry (Lara et al., 1997; Krasnopolsky and Cruikshank, 1995). Indeed, photochemical haze was seen in Triton's atmosphere by Voyager 2 (Herbert and Sandel, 1991; Rages and Pollack, 1992). The most common photochemical products of the interaction between CH$_4$ and extreme-UV photons, cosmic rays, and charged particles from Neptune's magnetosphere are acetylene (C$_2$H$_2$), ethylene (C$_2$H$_4$), and ethane (Krasnopolsky and Cruikshank, 1995; Moore and Hudson, 2003). Higher order hydrocarbons may also be created in much smaller quantities. Photochemistry occurs primarily in Triton's atmosphere, as demonstrated by a calculation of the flux of Lyman-$\alpha$ photons reaching the surface. The optical depth for Lyman-$\alpha$ photons in Triton's atmosphere, $\tau$, is the product of the UV cross-section of CH$_4$ at 120 nm (1.8$\times$10$^{-17}$ cm$^2$; Chen and Wu, 2004) and the column density of CH$_4$ (2.15$\times$10$^{18}$cm$^{-2}$; Lellouch et al., 2010). This results in $\tau$=38.7 and means that the flux of Lyman-$\alpha$ photons reaching Triton's surface is so exceedingly small as to be negligible. Photochemical reactions involving CH$_4$ occur exculsively in Triton's atmosphere.

\begin{figure}[h!]
\begin{center}
\includegraphics[scale=0.52,trim=0cm 2cm 1.0cm 3.0cm,clip=true]{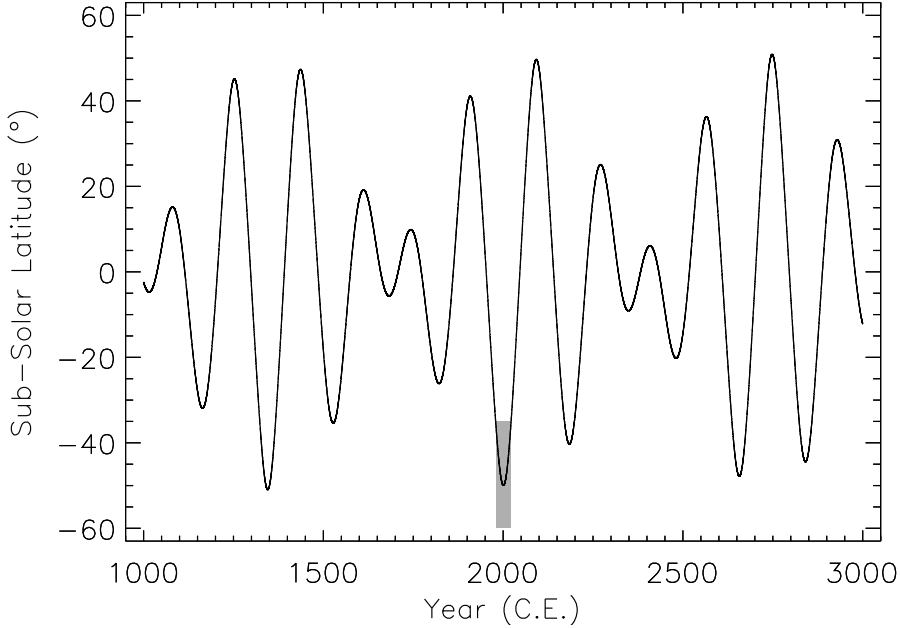}
\caption{Sub-solar latitude on Triton between 1000 and 3000 C.E. (data from JPL HORIZONS). The combination of Neptune's obliquity (30$^{\circ}$), the inclination of Triton's orbit (20$^{\circ}$), and the rapid precession of Triton's orbital node (637$\pm$40 years) contribute to Triton's unique seasons (Trafton, 1984). This results in a beat pattern between the precession period and the 165-year orbit of Neptune. The shaded region covers the years 1980-2020 and is presented in more detail in Fig. 2.}
\end{center}
\end{figure}

\begin{figure}[h!]
\begin{center}
\includegraphics[scale=0.52,trim=0cm 2cm 1.5cm 3cm,clip=true]{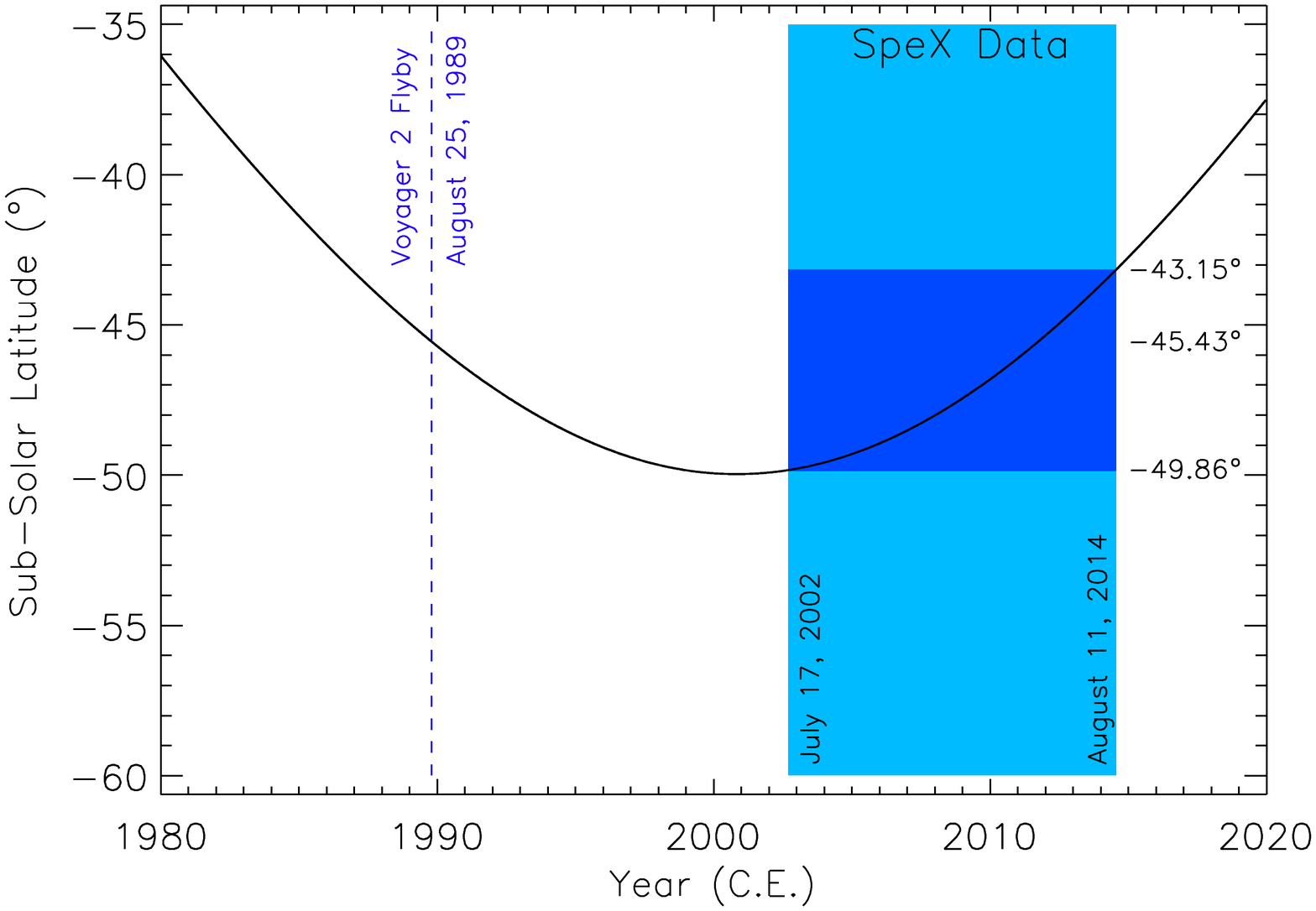}
\caption{Sub-solar latitude on Triton between 1980 and 2020 C.E. (data from JPL HORIZONS). The maximum southern excursion of the sub-solar latitude was approximately -50$^{\circ}$ and occurred in 2000. The Voyager 2 flyby took place when the sub-solar latitude was -45.43$^{\circ}$. The lighter region denotes the time baseline of the IRTF/SpeX observations analyzed in this paper. The darker region denotes the range of sub-solar latitudes observed. The sub-solar latitude has monotonically increased over the period of our observations.}
\end{center}
\end{figure}

\indent Ethane, a photochemical product produced in large quantities (Krasnopolsky and Cruikshank, 1995), was cited as a possible constituent ice on Triton's surface by DeMeo et al. (2010). Theoretically, after forming in the atmosphere, non-volatile ethane would precipitate onto Triton's surface and accumulate over time. Once a 1 $\mu$m spherical grain forms, it precipitates out of the atmosphere at the Stokes' velocity (Mark Bullock, private communication):
\begin{equation}
v=\frac{2gR^2(\rho -\rho_{atm})}{9\mu},
\end{equation}
where $g$ is the gravitational acceleration on Triton (77.9 cm s$^{-2}$), $R$ is the radius of the ethane grains, $\rho$ is the density of ethane (0.713 g cm$^{-3}$; Barth and Toon, 2003), $\rho_{atm}$ is the density of the atmosphere, and $\mu$ is the dynamic viscosity of the atmosphere. The number density of the dominant component of Triton's atmosphere, N$_2$, is 10$^{16}$ cm$^{-3}$ at the surface (Herbert and Sandel, 1991), and we used this to calculate a mass density of 4.65$\times$10$^{-7}$ g cm$^{-3}$. This value is negligible compared to the density of solid ethane, so it was ignored in the calculation. We calculate the dynamic viscosity using Sutherland's formula (Eq. (2)). This gives a lower limit on the viscosity and therefore an upper limit to the Stokes' velocity since the effects of turbulence and eddies are ignored. The dynamic viscosity ($\mu$) of an N$_2$ atmosphere at $T$=50 K (temperature of Triton's troposphere; Tyler et al., 1989) is:
\begin{equation}
\mu=\mu_0\frac{T_0+C}{T+C}\left(\frac{T}{T_0}\right)^{3/2},
\end{equation}
where $T_0$ is a reference temperature (300.55 K for N$_2$), $\mu_0$ is the dynamic viscosity at the reference temperature (0.0002 g cm$^{-1}$ s$^{-1}$ for N$_2$), and $C$ is Sutherland's constant for the gas in question (111 K for N$_2$).  We calculate $\mu$=0.00003 g cm$^{-1}$ s$^{-1}$. Combining all this information yielded a descent velocity of $\sim$0.004 cm s$^{-1}$ for a 1 $\mu$m ethane grain. Ethane more than likely does not form perfectly spherical grains, but instead takes the form of fractal aggregates; we assume spheres for ease of calculation, so the descent velocity quoted is an upper limit. Horizontal winds at altitudes below 8 km on Triton are $\sim$500 cm s$^{-1}$ (Ingersoll, 1990). This means that within one equatorial circumnavigation of Triton, an ethane grain would descend, at most, on the order of 100 m from its altitude of origin. Based on this calculation, we would expect ethane to precipitate uniformly onto Triton's surface independent of where it formed in the atmosphere. With an accumulation rate of 28 g cm$^{-2}$ Gyr$^{-1}$ (Krasnopolsky and Cruikshank, 1995), we calculate that a total of 180 cm of ethane has precipitated uniformly onto the surface of Triton over the age of the solar system. This assumes that Triton's atmosphere was in a steady state for the past 4.56 Gyr, so the calculated value is a rough upper limit.\\
\indent Ethane ice was previously detected on Pluto, an icy, outer solar system body similar in size to Triton (DeMeo et al., 2010; Holler et al., 2014). DeMeo et al. (2010) also made a tentative detection of ethane on Triton. In this work we study the longitudinal variability of the 2.405 $\mu$m absorption feature and test whether $^{13}$CO or ethane is responsible for its presence. The 2.405 $\mu$m feature is interesting because it could be due to ethane, $^{13}$CO, or both (Cruikshank et al., 2006). DeMeo et al. (2010) find that the 2.405 $\mu$m feature in Triton's spectrum is too strong to be due to ethane absorption alone; $^{13}$CO is therefore partially responsible for the observed depth. They also suggest the same for Pluto, presenting the possibility of a systematic error such as inaccurate optical constants. Work by Cruikshank et al. (2006) suggests that the 2.405 $\mu$m band in the spectra of Pluto and Triton is due to ethane and not $^{13}$CO.\\
\indent Sufficient solar illumination and the presence of strong vibrational transitions of N$_2$, CH$_4$, CO, CO$_2$, and H$_2$O make the near-infrared (0.7-2.5 $\mu$m) ideal for determining longitudinal distributions and for quantifying temporal evolution of these ices on the surface of Triton. We obtained reflectance spectra of Triton between 2002-2014 using the same telescope and instrument, allowing for easy comparison and combination of data in an effort to describe the effects of volatile transport on Triton's surface in both space and time. We also investigated the spatial and temporal variability of the absorption feature at 2.405 $\mu$m using this large data set.}

\section{Observations \& Reduction}
\onehalfspacing{Triton was observed from 2002-2014 using the SpeX infrared spectrograph at NASA's 3-meter Infrared Telescope Facility (IRTF) atop Mauna Kea (Rayner et al., 1998, 2003). During this period, 63 nights of usable reflectance spectra were obtained with observational circumstances found in Table 1 (information for 2002/07/17 to 2009/07/21 adapted from Grundy et al., 2010). Text files including the spectra for all good nights can be found at \textit{www2.lowell.edu/users/grundy/abstracts/2010.Triton.html} and as supplementary material online. Spectra were obtained in the short cross-dispersed mode using a 0.3"$\times$15'' slit. We aimed for a rotation angle close to the paralactic angle while avoiding angles within 15$^{\circ}$ of the imaginary line connecting Neptune and Triton. Following this procedure significantly reduced flux contamination from Neptune. All spectra obtained prior to 2014 were resampled onto a consistent wavelength grid covering 0.67-2.55 $\mu$m at $\lambda/\Delta\lambda$$\sim$2000 to match observations made with SpeX following its 2014 upgrade.\\
\indent Triton spectra were obtained using an ABBA slit dither pattern, with 2-minute integrations at each A and B position. Observations of Triton were interspersed with observations of a nearby solar analog star roughly once an hour. Solar analog spectra were also obtained following an ABBA dither pattern, with integration times dependent on the star. A solar analog requiring a large slew away from Neptune's position would potentially result in a less-than-optimal airmass correction and cause instrumental flexure, so three different stars were used between 2002 and 2014. New solar analogs were found when Neptune's motion across the sky produced too large a separation from the previous solar analog. From 2002-2007, HD 202282 (spectral type G3 V; Houk and Smith-Moore, 1988) was used as the solar analog. From 2007-2014, BS 8283 (binary with spectral types G0 V and G1 IV; Neckel, 1986; Pourbaix et al., 2004) was used as the solar analog. This binary star was checked against HD 202282 in 2007 to ensure that it could serve as the solar analog. Starting in 2014, HD 215295 (spectral type G2/3 V; Houk and Swift, 1999) was used as the solar analog.\\
\indent An optimal extraction algorithm was used to reduce the Triton spectra (Horne, 1986). Wavelength calibration was performed using spectra of argon lamps contained within the SpeX instrument as well as sky emission lines in the Triton spectra themselves. The use of AB pairs eliminated sky emission lines. Airmass-corrected solar analog spectra were matched to Triton spectra with similar airmasses from the same night. Dividing the Triton spectra by the matching solar analog spectra removed solar absorption lines and corrected for telluric absorption. Some telluric absorption persisted because it is not possible for the Triton and solar analog spectra to be obtained simultaneously at the same airmass. All spectra were then normalized. A grand average spectrum of all 63 individual nightly spectra was constructed by performing a weighted average within each wavelength bin (Fig. 3). Geometric albedos were not obtainable because the slit did not admit all light from the target object.}

\begin{figure}[h!]
\begin{center}
\includegraphics[scale=0.61,trim=0cm 1.8cm 1.5cm 3.25cm,clip=true]{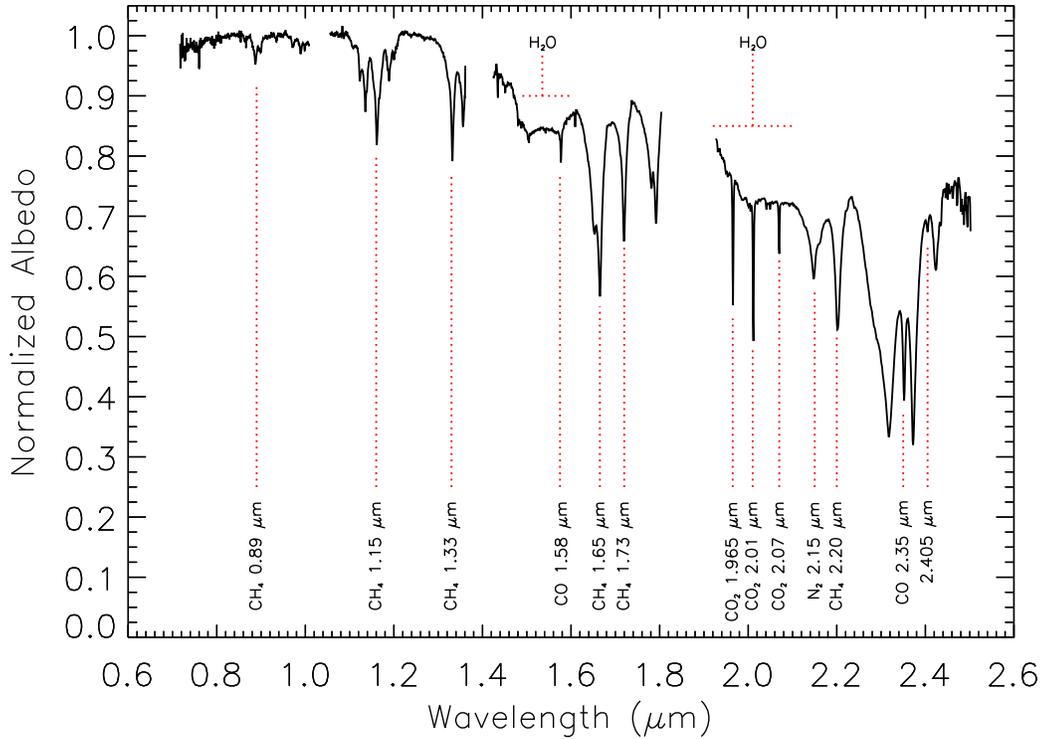}
\caption{Combined Triton grand average spectrum of the 63 individual spectra obtained between 2002 and 2014. Regions affected by significant telluric absorption were removed. The SpeX instrument currently covers 0.67-2.55 $\mu$m at a spectral resolution of $\sim$2000. All 63 spectra were resampled onto a consistent wavelength grid before being combined. The absorption bands considered in the analysis are identified and marked with dotted lines.}
\end{center}
\end{figure}

\begin{table}[h!]
\begin{center}
\textbf{Table 1}\\
Observational Circumstances\\
\begin{tabular}{cccccc}
\hline
UT date & Weather and & Sub-Earth & Sub-Earth & Phase & Total exp.\\
mean-time & \textit{H}-band seeing & Lon. ($^{\circ}$E) & Lat. ($^{\circ}$S) & angle ($^{\circ}$) & time (min)\\
\hline
2002/07/17 09:53 & Cirrus, 0.8" & 124.3 & 49.8 & 0.52 & 84\\
2002/07/18 09:36 & Thin cirrus, 0.8" & 184.8 & 49.8 & 0.49 & 64\\
2002/07/19 09:31 & Thin cirrus, 0.7" & 245.8 & 49.8 & 0.46 & 68\\
2002/07/20 09:27 & Thin cirrus, 0.8" & 306.9 & 49.8 & 0.42 & 68\\
2002/07/21 09:29 & Thin cirrus, 0.6" & 8.2 & 49.8 & 0.39 & 68\\
2002/07/22 09:21 & Clear, 0.7" & 69.1 & 49.8 & 0.36 & 84\\
2002/09/16 06:19 & Cirrus, 0.5" & 249.5 & 49.9 & 1.34 & 84\\
2002/10/03 06:32 & N/A, 0.6" & 211.0 & 49.9 & 1.67 & 48\\
2003/07/04 13:58 & Clear, 0.5" & 101.0 & 49.6 & 0.98 & 64\\
2003/08/10 10:09 & Some clouds, 0.5" & 196.4 & 49.7 & 0.19 & 58\\
2003/09/10 09:32 & Clear, 0.6" & 292.6 & 49.8 & 1.13 & 40\\
2003/10/16 05:51 & Patchy clouds, 0.5" & 327.5 & 49.8 & 1.80 & 88\\
2004/06/28 13:07 & Cirrus, 0.5" & 194.9 & 49.3 & 1.19 & 152\\
2004/08/12 12:19 & Clear, 0.7" & 67.7 & 49.5 & 0.21 & 52\\
2004/09/12 10:51 & Clear, 1.0" & 161.7 & 49.6 & 1.15 & 48\\
2004/10/21 06:13 & Partly cloudy, 0.8" & 18.0 & 49.6 & 1.84 & 52\\
2005/07/04 13:06 & Clear, 0.7" & 244.6 & 48.9 & 1.10 & 146\\
2005/09/19 08:02 & Clear, 0.5" & 265.2 & 49.3 & 1.26 & 68\\
2006/05/26 14:52 & Clear, 0.4" & 62.3 & 48.4 & 1.86 & 40\\
2006/06/26 13:39 & Cirrus, 0.4" & 157.9 & 48.5 & 1.37 & 128\\
2006/07/26 11:01 & Clear, 0.9" & 188.0 & 48.6 & 0.53 & 72\\
2006/08/30 09:14 & Clear, 0.7" & 166.0 & 48.8 & 0.62 & 48\\
2006/10/28 05:59 & Thin cirrus, 0.6" & 170.3 & 49.0 & 1.85 & 68\\
2007/06/21 13:15 & Clear, 0.8" & 252.6 & 47.9 & 1.53 & 136\\
2007/06/22 13:10 & Thin cirrus, 0.6" & 313.7 & 47.9 & 1.51 & 160\\
2007/06/23 13:18 & Clear, 0.6" & 15.2 & 47.9 & 1.49 & 128\\
2007/06/24 14:06 & Clear, 0.7" & 78.5 & 47.9 & 1.47 & 88\\
2007/06/25 13:13 & Clear, 0.5" & 137.5 & 47.9 & 1.45 & 132\\
2008/07/14 10:57 & Clear, 0.5" & 318.3 & 47.4 & 1.01 & 72\\
2008/09/01 07:02 & Clear, 0.6" & 68.1 & 47.8 & 0.55 & 56\\
2008/09/28 09:55 & Partly cloudy, 0.5" & 288.5 & 48.0 & 1.32 & 60\\
2009/06/22 14:35 & Clear, 0.7" & 102.2 & 46.6 & 1.59 & 32\\
2009/06/24 14:25 & Clear, 0.6" & 224.3 & 46.6 & 1.56 & 80\\
2009/07/21 12:07 & Clear, 0.5" & 71.7 & 46.8 & 0.88 & 64\\
2009/08/13 12:07 & Mostly clear, 0.6" & 39.8 & 47.0 & 0.15 & 56\\
2009/08/31 11:07 & N/A, 0.5" & 60.0 & 47.2 & 0.44 & 42\\
2010/08/07 11:08 & N/A, 0.7" & 71.9 & 46.3 & 0.43 & 56\\
2010/08/30 09:55 & Clear, 0.9" & 36.9 & 46.5 & 0.33 & 80\\
2011/06/11 15:01 & Clear, 0.5" & 233.0 & 45.1 & 1.83 & 28\\
2011/06/12 14:23 & Cirrus, 0.4" & 292.7 & 45.1 & 1.82 & 86\\
2011/06/20 14:40 & Clear, 1.0" & 63.4 & 45.1 & 1.72 & 50\\
2011/06/28 14:31 & Cirrus, 0.7" & 193.0 & 45.2 & 1.58 & 60\\
2011/08/06 12:04 & N/A, 0.6" & 54.8 & 45.5 & 0.55 & 60\\
\hline
\end{tabular}
\end{center}
\end{table}

\clearpage

\begin{table}[h!]
\begin{center}
\textbf{Table 1 (cont.)}\\
Observational Circumstances\\
\begin{tabular}{cccccc}
\hline
UT date & Weather and & Sub-Earth & Sub-Earth & Phase & Total exp.\\
mean-time & \textit{H}-band seeing & Lon. ($^{\circ}$E) & Lat. ($^{\circ}$S) & angle ($^{\circ}$) & time (min)\\
\hline
2011/08/23 09:08 & N/A, 0.5" & 8.1 & 45.6 & 0.02 & 64\\
2011/09/18 09:42 & N/A, 0.5" & 161.3 & 45.9 & 0.84 & 56\\
2011/11/13 06:07 & Mostly clear, 0.9" & 341.4 & 46.2 & 1.87 & 144\\
2012/06/25 14:19 & Clear, 0.7" & 111.6 & 44.3 & 1.66 & 80\\
2012/07/18 14:43 & Clear, 0.7" & 81.1 & 44.5 & 1.15 & 66\\
2012/09/02 07:01 & N/A, 0.7" & 357.9 & 44.9 & 0.29 & 72\\
2012/09/05 09:44 & Cirrus, 0.7" & 188.5 & 45.0 & 0.39 & 284\\
2012/09/08 08:41 & Cirrus, 0.7" & 9.5 & 45.0 & 0.49 & 200\\
2012/09/09 09:52 & Clear, 0.7" & 73.7 & 45.0 & 0.52 & 224\\
2012/09/10 09:51 & Clear, 0.6" & 134.9 & 45.0 & 0.55 & 228\\
2012/09/12 10:03 & Clear, 0.7" & 257.9 & 45.1 & 0.61 & 232\\
2012/09/13 08:54 & Clear, 0.7" & 316.1 & 45.1 & 0.64 & 248\\
2013/07/23 15:11 & N/A, 0.7" & 70.1 & 43.6 & 1.09 & 36\\
2013/07/25 14:55 & Clear, 0.9" & 191.9 & 43.6 & 1.03 & 48\\
2013/07/28 14:57 & N/A, 0.6" & 15.7 & 43.6 & 0.95 & 56\\
2014/08/04 13:49 & Mostly clear, 1.0'' & 122.9 & 42.8 & 0.82 & 135\\
2014/08/05 13:19 & Clear, 0.7'' & 182.9 & 42.8 & 0.79 & 155\\
2014/08/06 13:10 & Clear, 0.5'' & 243.7 & 42.8 & 0.76 & 175\\
2014/08/10 12:57 & Clear, 0.7'' & 128.1 & 42.8 & 0.63 & 167\\
2014/08/11 12:59 & Clear, 0.3'' & 189.4 & 42.9 & 0.60 & 171\\
\hline
\end{tabular}
\end{center}
\end{table}

\section{Analysis}
\subsection{N$_2$, CO, CH$_4$, CO$_2$, and the 2.405 $\mu$m band}
\onehalfspace{We calculated the integrated band areas (identical to an equivalent width calculation for a normalized spectrum) of the absorption bands presented in Table 2 in order to quantify N$_2$, CO, CH$_4$, CO$_2$, and 2.405 $\mu$m band variations with longitude and time on the surface of Triton. As a first step, we determined the wavelength range of three separate regions around each feature: pre-absorption continuum, band, and post-absorption continuum. These wavelength ranges are presented for each absorption band in Table 2. A least squares linear fit was performed across the band on the points within the pre- and post-continuum regions (in this context, continuum does not refer to a region completely devoid of absorption). After the spectrum was normalized by the best-fit line, the integrated band area was calculated via numerical integration. The errors on the integrated band areas were computed in an identical manner to those in Grundy et al. (2010).\\
\indent The integrated band area of each night-averaged spectrum was plotted against its corresponding sub-observer longitude. A sinusoid of the form $y=A$sin$(x+B)+C$, where $y$ is integrated band area (in $\mu$m) and $x$ is longitude, was robustly fit to the longitudinal (zonal) distribution using a non-linear least squares algorithm. The data were initially fit to a sinusoid and the residuals and standard deviation of the residuals were calculated; data points with residuals greater than 3-$\sigma$ were removed and another sinusoidal fit performed (Buie and Bus, 1992). This robust-fitting process was repeated a total of three times. The period was fixed at 2$\pi$, representing one rotation of Triton. The reduced $\chi^2$ was calculated for the sinusoidal fit and a horizontal line placed at the mean value of $C$ (to model zero variability over one Triton rotation). The number of degrees of freedom was 60 for the sinusoidal fit and 62 for the no-variability model. The longitude of peak absorption, peak-to-peak amplitude (twice the amplitude divided by the minimum), mean ($C$), and reduced $\chi^2$ values of the sinusoidal and no-variability models for each absorption band are presented in Table 3.\\
\indent To quantify temporal change, the integrated band area for each night-averaged spectrum was plotted against the corresponding Julian date. The observations from each night were made over a small range of sub-observer longitudes, so to disentangle the effects of longitudinal and temporal variability the value of the sinusoidal fit at the mean longitude for each night was subtracted off. The mean was added back on to maintain a similar mean value as the longitudinal fit. We plotted these values against phase angle for each absorption band to determine if the resulting distributions had a slope or followed a non-linear functional form. When plotted against phase angle, all distributions were essentially reasonable scatter about a mean, so phase angle effects were not considered any further in the analysis. A robust linear least squares fit was performed to determine if a particular ice species saw an increase or decrease in absorption between 2002 and 2014. The slope ($m$), slope detection level ($m/\sigma_m$), and $y$-intercept ($b$) for each absorption band are found in Table 4.\\
\indent We also investigated how combinations of CH$_4$ bands varied with longitude and time. In particular, we combined the weak CH$_4$ bands (0.89, 1.15, and 1.33 $\mu$m), strong CH$_4$ bands (1.65, 1.73, and 2.20 $\mu$m), and all CH$_4$ bands. Combination involved addition of the integrated band areas for the appropriate CH$_4$ bands on each night and propagation of errors. A combination of the bands comprising the CO$_2$ triplet (1.965, 2.01, and 2.07 $\mu$m) was constructed in an identical manner. A combination of the 1.58 and 2.35 $\mu$m CO bands was performed but the results did not add any new information. Results for the sinusoidal fit to longitude and the linear fit to time for each combination are found in Tables 3 and 4, respectively.}

\subsection{H$_2$O}
\onehalfspace{Analysis of the longitudinal distribution and temporal evolution of H$_2$O ice on Triton proceeded differently than for the other ices. As seen in Fig. 3, absorption due to H$_2$O is very broad and overlaps with absorption bands of CH$_4$, CO, and CO$_2$. This prevented the calculation of an integrated band area, so instead we calculated fractional band depth for H$_2$O. The fractional band depth was computed for each individual spectrum by dividing the mean value of the points in a relatively flat region of H$_2$O absorption (1.5099-1.5694 $\mu$m) by the mean value of the points in a suitable continuum region (1.2195-1.2693 $\mu$m), and subtracting this quantity from one. While the fractional band depth is an even cruder estimation of abundance than the integrated band area, it still provides information on how H$_2$O abundance differs with longitude and time. We aimed to determine differences and not absolute amounts of any species in this paper. The procedures described in the previous section were used to determine the longitudinal distribution and temporal evolution of H$_2$O ice on the surface of Triton. Information on the sinusoidal fit to longitude and the linear fit to time are found in Tables 3 and 4, respectively.}

\begin{table}[h!]
\begin{center}
\textbf{Table 2}\\
Continuum and Band Regions\\
\begin{tabular}{cccc}
\hline
Absorption band & Pre-absorption ($\mu$m) & Band ($\mu$m) & Post-absorption ($\mu$m)\\
\hline
N$_2$ (2.15 $\mu$m) & 2.0925-2.1164 & 2.1164-2.1739 & 2.1739-2.1838\\
CO (1.58 $\mu$m) & 1.5510-1.5701 & 1.5701-1.5859 & 1.5859-1.6047\\
CO (2.35 $\mu$m) & 2.3336-2.3453 & 2.3453-2.3592 & 2.3592-2.3678\\
CO$_2$ (1.965 $\mu$m) & 1.9564-1.9608 & 1.9608-1.9707 & 1.9707-1.9850\\
CO$_2$ (2.01 $\mu$m) & 1.9995-2.0068 & 2.0068-2.0178 & 2.0178-2.0381\\
CO$_2$ (2.07 $\mu$m) & 2.0539-2.0651 & 2.0651-2.0736 & 2.0736-2.0963\\
2.405 $\mu$m & 2.3970-2.4014 & 2.4014-2.4090 & 2.4090-2.4134\\
\textbf{Weak CH$_4$ bands:} & & &\\
CH$_4$ (0.89 $\mu$m) & 0.8595-0.8776 & 0.8776-0.9085 & 0.9085-0.9298\\
CH$_4$ (1.15 $\mu$m) & 1.0694-1.0995 & 1.0995-1.2096 & 1.2096-1.2296\\
CH$_4$ (1.33 $\mu$m) & 1.2402-1.2920 & 1.2920-1.3423 & 1.3423-1.3435\\
\textbf{Strong CH$_4$ bands:} & & &\\
CH$_4$ (1.65 $\mu$m) & 1.6040-1.6194 & 1.6194-1.6839 & 1.6839-1.6985\\
CH$_4$ (1.73 $\mu$m) & 1.6855-1.6970 & 1.6970-1.7368 & 1.7368-1.7439\\
CH$_4$ (2.20 $\mu$m) & 2.1739-2.1838 & 2.1838-2.2279 & 2.2279-2.2360\\
\hline
\end{tabular}
\end{center}
\end{table}

\clearpage

\begin{table}[h!]
\begin{center}
\textbf{Table 3}\\
Rotational Sinusoidal Fits\\
\begin{tabular}{cccccc}
\hline
Absorption & Peak & Peak-to-peak & Mean & \multicolumn{2}{c}{Reduced $\chi^2$}\\
band & long. ($^{\circ}$) & amp. (\%) & (10$^{-6}$ $\mu$m) & Sine & Line\\
\hline
N$_2$ (2.15 $\mu$m) & 35$\pm$1  & 68$\pm$1 & 3371.6$\pm$8.8 & 2.8 & 45\\
CO (1.58 $\mu$m) & 60$\pm$3 & 33$\pm$1 & 342.6$\pm$1.5 & 1.3 & 2.1\\
CO (2.35 $\mu$m) & 55$\pm$4 & 90$\pm$6 & 1016$\pm$13 & 0.9 & 3.7\\
CO$_2$ (1.965 $\mu$m) & 97$\pm$54 & 8$\pm$8 & 732$\pm$20 & 0.1 & 0.1\\
CO$_2$ (2.01 $\mu$m) & 124$\pm$83 & 4$\pm$7 & 846$\pm$20 & 0.3 & 0.3\\
CO$_2$ (2.07 $\mu$m) & 236$\pm$12 & 15$\pm$3 & 297.6$\pm$2.9 & 0.2 & 0.2\\
CO$_2$ (All) & 128$\pm$61 & 4$\pm$5 & 1877$\pm$31 & 0.3 & 0.3\\
H$_2$O\renewcommand{\thefootnote}{\fnsymbol{footnote}}\footnote{The fractional band depth was calculated for the H$_2$O absorption instead of the integrated band area. Therefore the mean value and uncertainty on the mean value are reported as unitless fractions.} & 228$\pm$100 & 1$\pm$1 & 0.15938$\pm$0.00085 & 1.5 & 1.5\\
2.405 $\mu$m & 152$\pm$44 & 30$\pm$25 & 95.3$\pm$7.1 & 1.1 & 1.1\\
CH$_4$ (All Avg.) & 296$\pm$1 & 22$\pm$1 & 24144.8$\pm$3.2 & 6.3 & 48\\
\textbf{Weak CH$_4$ bands:} & & & & & \\
CH$_4$ (Weak Avg.) & 285$\pm$1 & 36$\pm$1 & 8620.4$\pm$1.7 & 15 & 58\\
CH$_4$ (0.89 $\mu$m) & 304$\pm$1 & 37$\pm$1 & 566.3$\pm$1.7 & 3.2 & 5.0\\
CH$_4$ (1.15 $\mu$m) & 284$\pm$1 & 37$\pm$1 & 6140.2$\pm$4.3 & 19 & 54\\
CH$_4$ (1.33 $\mu$m) & 284$\pm$2 & 32$\pm$1 & 1901.9$\pm$5.6 & 5.4 & 14\\
\textbf{Strong CH$_4$ bands:} & & & & & \\
CH$_4$ (Strong Avg.) & 307$\pm$2 & 16$\pm$1 & 15514$\pm$28 & 3.2 & 16.6\\
CH$_4$ (1.65 $\mu$m) & 313$\pm$2 & 15$\pm$1 & 8068$\pm$14 & 2.4 & 14\\
CH$_4$ (1.73 $\mu$m) & 315$\pm$4 & 19$\pm$1 & 3050$\pm$13 & 1.4 & 4.5\\
CH$_4$ (2.20 $\mu$m) & 289$\pm$1 & 17$\pm$1 & 4397.70$\pm$0.60 & 1.9 & 4.3\\
\hline
\end{tabular}
\end{center}
\end{table}

\clearpage

\begin{table}[h!]
\begin{center}
\textbf{Table 4}\\
Secular Linear Fits\\
\begin{tabular}{cccccc}
\hline
Absorption & $m\pm\sigma_m$ & $m/\sigma_m$ & $b\pm\sigma_b$ &  \multicolumn{2}{c}{Reduced $\chi^2$}\\
band & (10$^{-6}$ $\mu$m yr$^{-1}$) & & (10$^{-6}$ $\mu$m) & \textbar $m$\textbar$>$0 & $m$=0\\
\hline
N$_2$ (2.15 $\mu$m) & 11.4$\pm$3.2 & 3.6 & 3373$\pm$12 & 2.6 & 2.8\\
CO (1.58 $\mu$m) & 4.6$\pm$1.2 & 4.0 & 342.0$\pm$4.8 & 1.0 & 1.3\\
CO (2.35 $\mu$m) & -1.4$\pm$4.5 & 0.3 & 1015$\pm$18 & 0.9 & 0.9\\
CO$_2$ (1.965 $\mu$m) & 2.6$\pm$4.6 & 0.6 & 732$\pm$19 & 0.1 & 0.1\\
CO$_2$ (2.01 $\mu$m) & 6.1$\pm$5.0 & 1.2 & 848$\pm$21 & 0.3 & 0.3\\
CO$_2$ (2.07 $\mu$m) & 1.0$\pm$2.4 & 0.4 & 297.5$\pm$9.7 & 0.2 & 0.2\\
CO$_2$ (All) & 9.6$\pm$7.2 & 1.3 & 1879$\pm$30 & 2.0 & 2.1\\
H$_2$O\renewcommand{\thefootnote}{\fnsymbol{footnote}}\footnote{The fractional band depth was calculated for the H$_2$O absorption instead of the integrated band area. The slope and the uncertainty on the slope are instead reported in units of fractional band depth per year. The $y$-intercept and the uncertainty on the $y$-intercept are instead reported as unitless fractions.} & 0.0$\pm$0.0 & 0.2 & 0.15938$\pm$0.00084 & 1.5 & 1.5\\
2.405 $\mu$m & -2.4$\pm$1.8 & 1.4 & 94.4$\pm$7.3 & 1.1 & 1.1\\
CH$_4$ (All) & 30.3$\pm$7.9 & 3.8 & 24154$\pm$32 & 6.0 & 6.1\\
\textbf{Weak CH$_4$ bands:} & & & & & \\
CH$_4$ (Weak) & -32.1$\pm$4.3 & 7.5 & 8607$\pm$18 & 14 & 15\\
CH$_4$ (0.89 $\mu$m) & -3.4$\pm$1.3 & 2.6 & 566.1$\pm$5.5 & 3.0 & 3.3\\
CH$_4$ (1.15 $\mu$m) & -25.4$\pm$3.5 & 7.3 & 6131$\pm$14 & 18 & 19\\
CH$_4$ (1.33 $\mu$m) & -3.7$\pm$2.0 & 1.9 & 1899.5$\pm$8.0 & 5.3 & 5.3\\
\textbf{Strong CH$_4$ bands:} & & & & & \\
CH$_4$ (Strong) & 63.7$\pm$6.6 & 9.7 & 15529$\pm$27 & 1.7 & 3.1\\
CH$_4$ (1.65 $\mu$m) & 23.1$\pm$3.4 & 6.8 & 8072$\pm$14 & 1.6 & 2.3\\
CH$_4$ (1.73 $\mu$m) & 14.8$\pm$3.1 & 4.7 & 3053$\pm$23 & 1.0 & 1.4\\
CH$_4$ (2.20 $\mu$m) & 26.4$\pm$4.7 & 5.6 & 4407$\pm$19 & 1.3 & 1.8\\
\hline
\end{tabular}
\end{center}
\end{table}

\section{Results \& Discussion}
\onehalfspace{The longitudinal (zonal) distributions of N$_2$, CO, CH$_4$ (all bands), CO$_2$ (all bands), H$_2$O, and the 2.405 $\mu$m band are presented in Fig. 4-7 and 9-11. A discussion of temporal variability is found later in this section. Plots presenting absorption as a function of time are provided for CH$_4$ (strong), N$_2$, CO (2.35 $\mu$m), and the CO$_2$ triplet in Fig. 13-16.\\
\indent The two species with the largest peak-to-peak amplitudes are the volatiles N$_2$ (Fig. 4) and CO (1.58 $\mu$m, Fig. 5; 2.35 $\mu$m, Fig. 6), with values of 68$\pm$1\%, 33$\pm$1\%, and 90$\pm$6\%, respectively. The longitudes of peak absorption are 35$\pm$1$^{\circ}$, 60$\pm$1$^{\circ}$, and 55$\pm$4$^{\circ}$, respectively. The phase difference between N$_2$ and the 1.58 $\mu$m CO band is 25$\pm$1$^{\circ}$; between N$_2$ and the 2.35 $\mu$m CO band it is 20$\pm$4$^{\circ}$; and between the two CO bands it is 5$\pm$5$^{\circ}$. The longitudinal distributions of N$_2$ and CO are nearly in-phase, suggesting these species migrate together across the surface of Triton. The large peak-to-peak amplitude as well as the relative and absolute longitudes of peak absorption of these species agree with previous work performed on a smaller SpeX data set in Grundy et al. (2010). Both ices have similar, non-negligible sublimation pressures (Fray and Schmitt, 2009) and are fully miscible in one another (Vetter et al., 2007). However, note that the distributions are not completely in-phase, even when accounting for the uncertainty. The 1.58 $\mu$m CO band is found within a region of H$_2$O absorption, but as described below, H$_2$O ice shows negligible variability over one Triton rotation. The 2.35 $\mu$m CO band is in between two strong CH$_4$ bands at 2.32 and 2.38 $\mu$m. However, the fact that the two CO bands are fully in-phase suggests that our choices of continua regions for each band eliminated any effect from the surrounding bands. Therefore, the peak longitudes of the CO bands are accurate and the shift with respect to N$_2$ is real. An explanation for this phenomenon is not immediately evident.\\
\indent Triton's spectrum is dominated by CH$_4$ absorption bands. We present here only the longitudinal variation of the sum of all CH$_4$ bands (Fig. 7) since the individual CH$_4$ bands all have roughly the same distribution. CH$_4$ shows moderate variability over one Triton rotation with a peak-to-peak amplitude of 22$\pm$1\%. The individual CH$_4$ bands, and thus the groupings of bands, all have peak longitudes in the region between $\sim$285-315$^{\circ}$ (Fig. 8), which is in agreement with the results from Grundy et al. (2010). These peak longitudes are found on the opposite side of the sub-Neptune hemisphere from those of N$_2$ and the two CO bands (phase differences of 99$\pm$1$^{\circ}$ from N$_2$, 124$\pm$3$^{\circ}$ from the 1.58 $\mu$m CO band, and 115$\pm$4$^{\circ}$ from the 2.35 $\mu$m CO band). While only small amounts of CH$_4$ can be diluted in N$_2$, and vice versa (Prokhvatilov and Yantsevich, 1983), there is no reason to believe that the N$_2$ and CH$_4$ distributions should be so far out of phase. In fact, it is possible for N$_2$:CH$_4$ and CH$_4$:N$_2$ mixtures to spatially coexist (Tegler et al., 2012; Protopapa et al., 2015). The fact that the two distributions are so far out of phase is unexpected. A similar pattern for the N$_2$, CO, and CH$_4$ distributions are seen on Pluto (Grundy et al., 2013). This similarity may imply more than mere coincidence and requires further study.\\
\indent The non-volatile ices CO$_2$ and H$_2$O show very little variation over one Triton rotation and have highly uncertain longitudes of peak absorption. We present the longitudinal variation of the sum of the CO$_2$ triplet (Fig. 9) rather than the individual CO$_2$ bands because the distributions are similar and the sum has smaller error bars. The peak longitude for the combination of CO$_2$ bands is 128$\pm$61$^{\circ}$ and the peak-to-peak amplitude is 4$\pm$5\%. We calculated the sinusoidal fit to the H$_2$O fractional band depths (Fig. 10) and obtained a peak longitude of 228$\pm$100$^{\circ}$ and a peak-to-peak amplitude of 1$\pm$1\%. The longitudinal distributions for both species cannot be distinguished between a sinusoid and a horizontal line (Table 3), meaning that their distributions across the surface of Triton are uniform. This is the same conclusion reached by Grundy et al. (2010) for the non-volatile ices. We note that the peak longitude (152$\pm$44$^{\circ}$), low amplitude, and reduced $\chi^2$ values for the 2.405 $\mu$m band suggest a uniform surface distribution as well.\\
\indent We investigated what effects, if any, the two CH$_4$ absorption bands at 2.38 and 2.43 $\mu$m had on the the longitudinal distribution of the 2.405 $\mu$m band. The effect of CH$_4$ absorption must be negligible since the 2.405 $\mu$m band distribution has such a low mean (0.0000953$\pm$0.0000071 $\mu$m) and a negligible amplitude (0.000012$\pm$0.000010 $\mu$m). The peak-to-peak amplitude for the 2.405 $\mu$m band is potentially large (30$\pm$25\%, rivaling the N$_2$ band), but this is misleading. The mean of the sinusoidal fit is very small and the amplitude is therefore large in comparison. However, the amplitude itself is low in an absolute sense. Contamination from more strongly variable CH$_4$ would be noticeable from its effects on both the mean and the amplitude of the sinusoidal fit. From this we conclude that our choice of continuum regions on either side of the 2.405 $\mu$m band effectively removed contamination from the adjacent CH$_4$ bands. The species responsible for the 2.405 $\mu$m band is distributed nearly uniformly across the surface of Triton.\\
\indent From the grand average spectrum (Fig. 3) we detect the 2.405 $\mu$m absorption band at the 4.5-$\sigma$ level. We compared the longitudinal distributions of CO (Fig. 5 \& 6) and the 2.405 $\mu$m band (Fig. 11) to determine the origin of the latter. The phase difference between the 2.405 $\mu$m band and the 1.58 $\mu$m CO band is 92$\pm$44$^{\circ}$. The phase difference between the 2.405 $\mu$m band and the 2.35 $\mu$m CO band is 97$\pm$44$^{\circ}$. Even if fractionation of CO were a factor on the surface of Triton, we would not expect the longitudinal distributions of $^{12}$CO and $^{13}$CO to be this far out of phase. Perhaps the quantity of $^{13}$CO present on Triton is too low to detect in reflectance spectra. Assuming that carbon on Triton matches solar composition, 1\% of all CO on Triton should be in the form of $^{13}$CO (Scott et al., 2006). Even if $^{13}$CO absorption at 2.405 $\mu$m were 1\% as strong as $^{12}$CO absorption at 2.35 $\mu$m (Fig. 3), the feature would be indistinguishable from noise. The longitudinal distribution of the 2.405 $\mu$m band also shares features with the distributions of non-volatile CO$_2$ and H$_2$O (Table 3): a poorly constrained peak longitude and low amplitude. This provides additional circumstantial evidence that the species responsible for the 2.405 $\mu$m band is also relatively non-volatile. Both $^{12}$CO and $^{13}$CO are volatile whereas ethane is non-volatile. This supports the interpretation that ethane is responsible for the absorption feature at 2.405 $\mu$m.\\
\indent A summary of the longitudinal distributions across Triton's surface is presented in Fig. 12. From this figure we note the large uncertainties on the peak longitudes for the non-volatile ices (CO$_2$, H$_2$O, ethane) and the smaller uncertainties on the peak longitudes for the volatile ices (N$_2$, CO, CH$_4$). The volatile ices also show larger variability across the surface of Triton, with the longitude of peak absorption likely corresponding to a large patch of a particular ice species. CO$_2$ and H$_2$O are uniformly distributed across the surface and act as the substrate upon which the volatile ices non-uniformly deposit (with ethane as a contaminant). One would expect the deposition of volatile ices to induce variability in the non-volatile ices, especially in regions of thick volatile coverage, yet non-volatile variability is negligible. The viewing geometry of Triton provides an explanation: Any latitude less than 90$^{\circ}$ minus the sub-solar latitude will be visible at all times. The simplest explanation for the observed variability of non-volatile ices is that they dominate the always visible south polar region. Therefore, the south polar region is relatively devoid of volatile ices. The volatile ices should be found farther north where parts of the surface rotate into and out of view.\\
\indent We support this claim with our analysis of the temporal evolution of the volatile and non-volatile ices. Statistically significant slopes ($\geq$3-$\sigma$) were found for the strong CH$_4$ band grouping, N$_2$, CO (1.58 $\mu$m), and CH$_4$ (all, weak band grouping, 1.15 $\mu$m, 1.65 $\mu$m, 1.73 $\mu$m, 2.20 $\mu$m). However, for all but the strong CH$_4$ band grouping (Fig. 13), the reduced $\chi^2$ values do not favor a sloped line over a horizontal line (Table 4). We find that a positively sloped line is statistically preferred for the strong CH$_4$ band grouping, meaning the quantity of CH$_4$ present on the visible portion of Triton increased over time. Figures 14-16 present the temporal distributions for N$_2$, CO (2.35 $\mu$m), and the CO$_2$ triplet. N$_2$ and CO are volatile ices that might be expected to show a similar change in time to volatile CH$_4$; the two CO bands show similar changes in time, so only one is presented. From our analysis, we conclude that the signal-to-noise ratios of the nightly spectra are not high enough to distinguish between a flat and non-zero slope for any ices other than CH$_4$. Grundy et al. (2010) found that the N$_2$ ice integrated band area showed a small decrease between 2000 and 2009 by considering two subsets of their data: 2000-2004 and 2005-2009 (see Fig. 11 of that paper). A slight downward shift in the longitudinal distribution was detected over this time period that they interpret as a change in the texture of the N$_2$ ice, not a change in areal coverage. A negligible change in N$_2$ absorption, as reported by Grundy et al. (2010), is consistent with the results of this work. Analysis of temporal variation for other species was not provided in their work.\\
\indent We considered the possibility that the observed increase in CH$_4$ absorption was due to volatile transport. Evidence for volatile transport on Triton dates to the 1950s, with conclusive evidence for surface color change between 1979 and the Voyager 2 flyby in 1989 (Smith et al., 1989; Buratti et al., 1994; Brown et al., 1995). The disk-averaged visible spectrum (0.3-0.8 $\mu$m) of Triton became significantly less red between 1979 and 1989 as the sub-solar latitude moved 10$^{\circ}$ south. This color change was interpreted as deposition of volatile ices onto older, redder regions of the surface. Bauer et al. (2010) present evidence for significant albedo changes across the surface of Triton between the Voyager 2 flyby in 1989 and 2005. The albedo of the equatorial regions and the sub-Neptune hemisphere increased and the albedo decreased in the anti-Neptune hemisphere over this period. We note that volatile ice absorptions peak on the sub-Neptune hemisphere (Fig. 12, this work). Fig. 2 from Bauer et al. (2010) shows the 2005 visible albedo map compared to the Voyager 2 albedo map. Shortly after the Voyager 2 flyby in 1989, many argued that volatile ices were found in the southern hemisphere due to the higher albedo (e.g., Stone and Miner, 1989; Moore and Spencer, 1990). This remains unconfirmed due to limited spectral information from Voyager 2.\\
\indent The increase in CH$_4$ absorption may instead be due to a change in viewing geometry. Our findings suggest that the southern latitudes are primarily denuded of volatile ices, revealing the underlying non-volatile substrate. We propose that volatile ices migrated during the southern summer to the northern hemisphere where they deposited in the region near the summer/winter terminator. As the sub-solar (and thus, the sub-observer) latitude moves northwards, we receive more direct reflection from these northern latitudes and less from the region surrounding the south pole. We define the south polar region to be the entire southern hemisphere due to its visual characteristics (Moore and Spencer, 1990). The northern temperate region is defined between 0$^{\circ}$ and the most northerly observable latitude. Imagine the sub-observer point at Triton's equator; the north pole and south pole are 90$^{\circ}$ away from the equator and just visible on the edge of the disk. For other values of the sub-observer latitude, the visible region will be bounded by the nearest pole at one extreme and the latitude 90$^{\circ}$ away on the other.\\
\indent At the start of the observations described in this work (July 2002), the sub-observer latitude was -49.86$^{\circ}$; on the night of the last observation in August 2014, it was -43.15$^{\circ}$. In both cases, the most southerly observable latitude was -90$^{\circ}$. In 2002, the most northerly observable latitude was 40.14$^{\circ}$; in 2014 it was 46.85$^{\circ}$. We developed code that calculated the projected solid angle of each small area of Triton's surface to determine the total projected area of a chosen range of latitudes. Then we divided this value by the total projected area (equal to $\pi$) to obtain the percentage of Triton's visible disk due to the chosen region. Due to longitudinal symmetry, this percentage applies at all rotational phases. See the Appendix for a more detailed mathematical description. The south polar region was responsible for 88.2\% of the projected area in 2002 and 84.2\% in 2014, a fractional decrease of 0.05. The northern temperate region was responsible for 11.8\% of the projected area in 2002 and 15.8\% in 2014, a fractional increase of 0.34.\\
\indent A significant increase in albedo near the longitude of peak absorption for CH$_4$ is seen in the 2005 map from Bauer et al. (2010). No such brightening is noticed at the longitudes of peak N$_2$ or CO absorption and the latitudes of increased albedo were visible in both 1989 and 2005 (this region was visible for the entire duration of our observations). These facts cannot be explained by a change in viewing geometry and a volatile transport explanation is not immediately evident. A variety of future observations and modeling of volatile transport on Triton will be necessary to explain the observed temporal changes.}

\clearpage

\begin{figure}[h!]
\begin{center}
\includegraphics[scale=0.55,trim=0cm 2.5cm 1.5cm 3.25cm,clip=true]{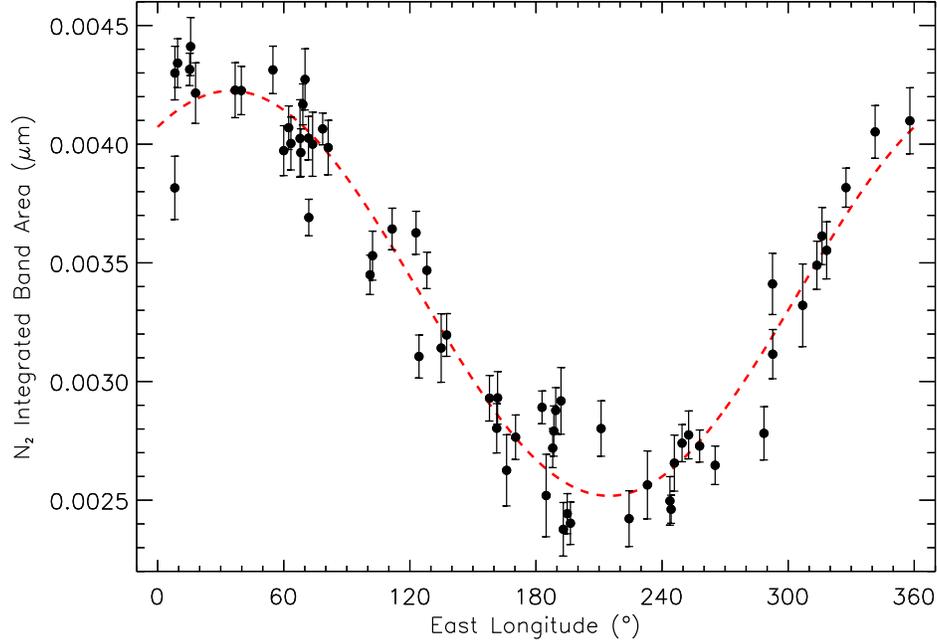}
\caption{Integrated band area of the 2.15 $\mu$m N$_2$ band plotted against sub-observer longitude. The dashed line is a robust sinusoidal fit to the points. N$_2$ ice shows significant variation over one Triton rotation.}
\end{center}
\end{figure}

\begin{figure}[h!]
\begin{center}
\includegraphics[scale=0.55,trim=0cm 2.5cm 1.5cm 3.25cm,clip=true]{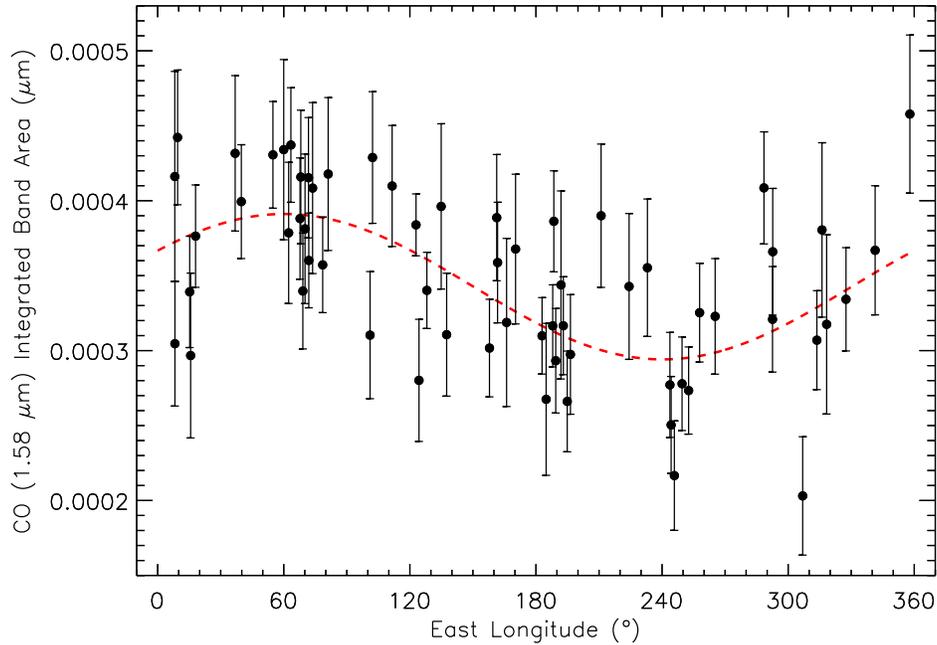}
\caption{Integrated band area of the 1.58 $\mu$m CO band plotted against sub-observer longitude. The dashed line is a robust sinusoidal fit to the points. The phase difference between the 1.58 $\mu$m and 2.35 $\mu$m CO bands is 5$\pm$5$^{\circ}$.}
\end{center}
\end{figure}

\begin{figure}[h!]
\begin{center}
\includegraphics[scale=0.55,trim=0cm 2.5cm 1.5cm 3.25cm,clip=true]{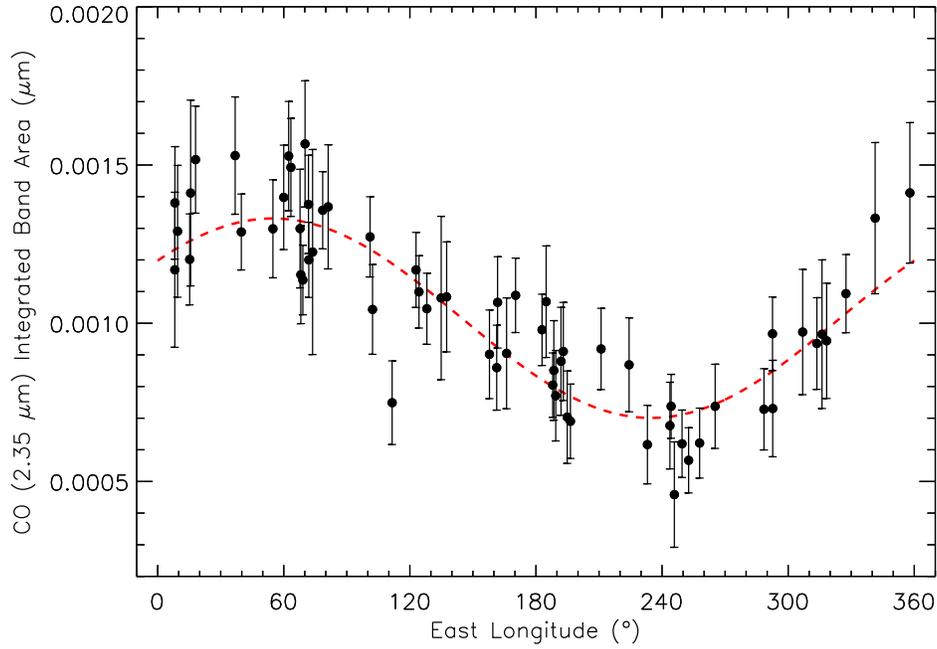}
\caption{Integrated band area of the 2.35 $\mu$m CO band plotted against sub-observer longitude. The dashed line is a robust sinusoidal fit to the points. CO  ice shows significant variation over one Triton rotation with a peak longitude similar to that of N$_2$ ice. This suggests that N$_2$ and CO migrate together as Triton's seasons change. The phase difference between the two CO bands is 5$\pm$5$^{\circ}$.}
\end{center}
\end{figure}

\begin{figure}[h!]
\begin{center}
\includegraphics[scale=0.55,trim=0cm 2.5cm 1.5cm 3.25cm,clip=true]{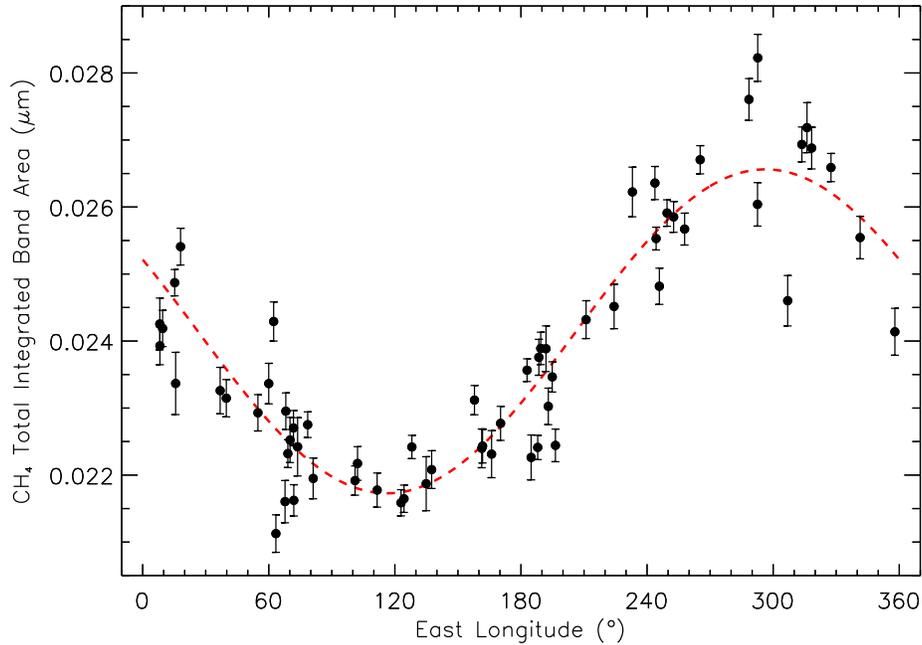}
\caption{Total integrated band area of all CH$_4$ bands plotted against sub-observer longitude. The dashed line is a robust sinusoidal fit to the points. CH$_4$ ice shows moderate variation over one Triton rotation.}
\end{center}
\end{figure}

\begin{figure}[h!]
\begin{center}
\includegraphics[scale=0.65,trim=0cm 5.2cm 1.5cm 3.25cm,clip=true]{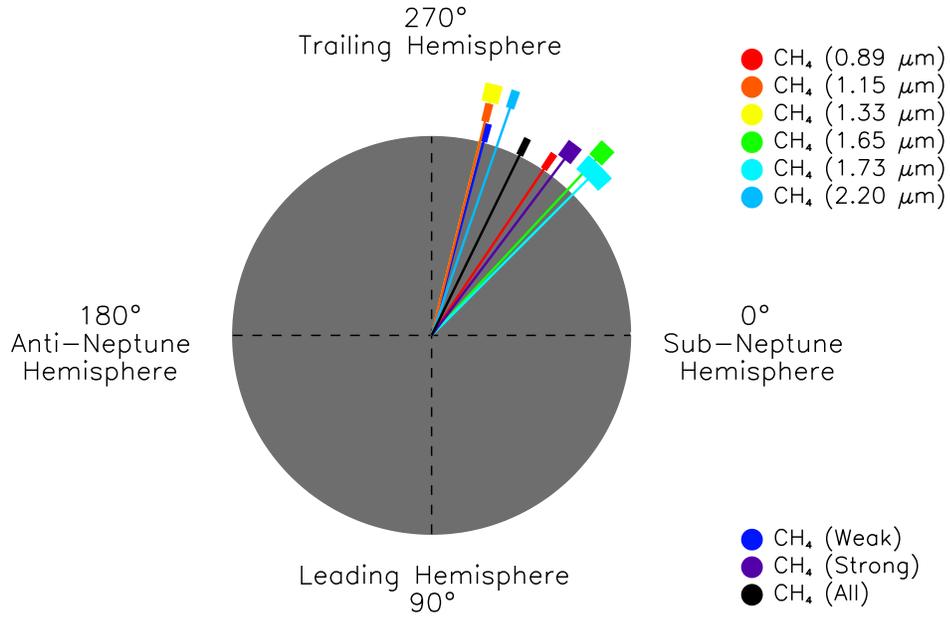}
\caption{Range of longitude of peak absorption (longitude $\pm$ uncertainty) for all CH$_4$ bands and combination of CH$_4$ bands. Bar sizes do not represent the extent of the ices but instead show the possible location of the peak longitude for that species. The 0.89, 1.15, and 1.33 $\mu$m bands comprise the weak CH$_4$ grouping. The 1.65, 1.73, and 2.20 $\mu$m bands comprise the strong CH$_4$ grouping.  Triton is viewed looking down on the south pole, with rotation and orbital motion counterclockwise.}
\end{center}
\end{figure}

\begin{figure}[h!]
\begin{center}
\includegraphics[scale=0.55,trim=0cm 2.5cm 1.5cm 3.25cm,clip=true]{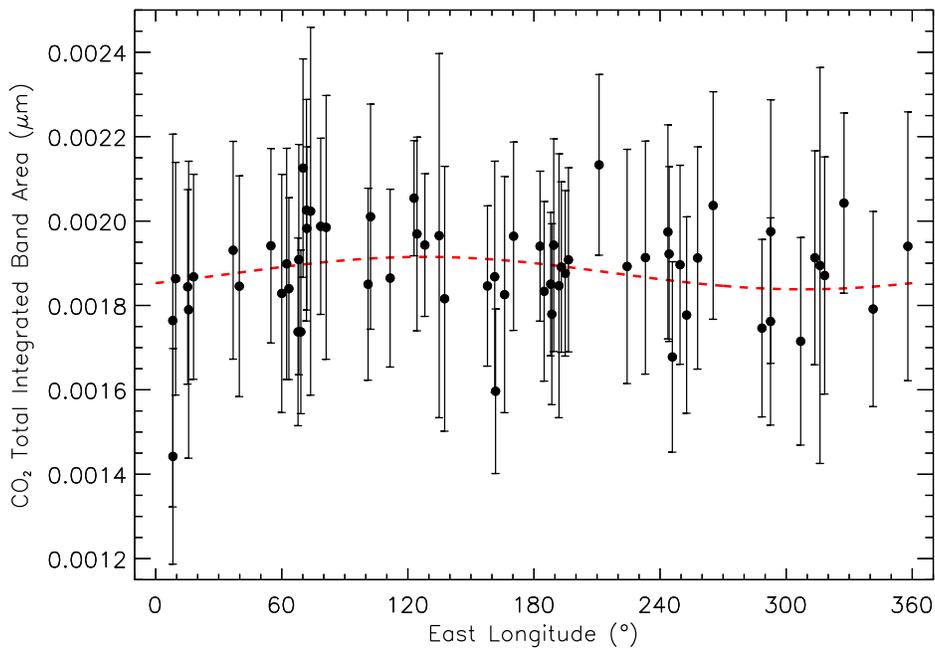}
\caption{Total integrated band area of all CO$_2$ bands plotted against sub-observer longitude. The dashed line is a robust sinusoidal fit to the points. CO$_2$ ice shows negligible variation over one Triton rotation. Additionally, the longitude of peak absorption for CO$_2$ is uncertain (Table 3).}
\end{center}
\end{figure}

\begin{figure}[h!]
\begin{center}
\includegraphics[scale=0.55,trim=0cm 2.5cm 1.5cm 3.25cm,clip=true]{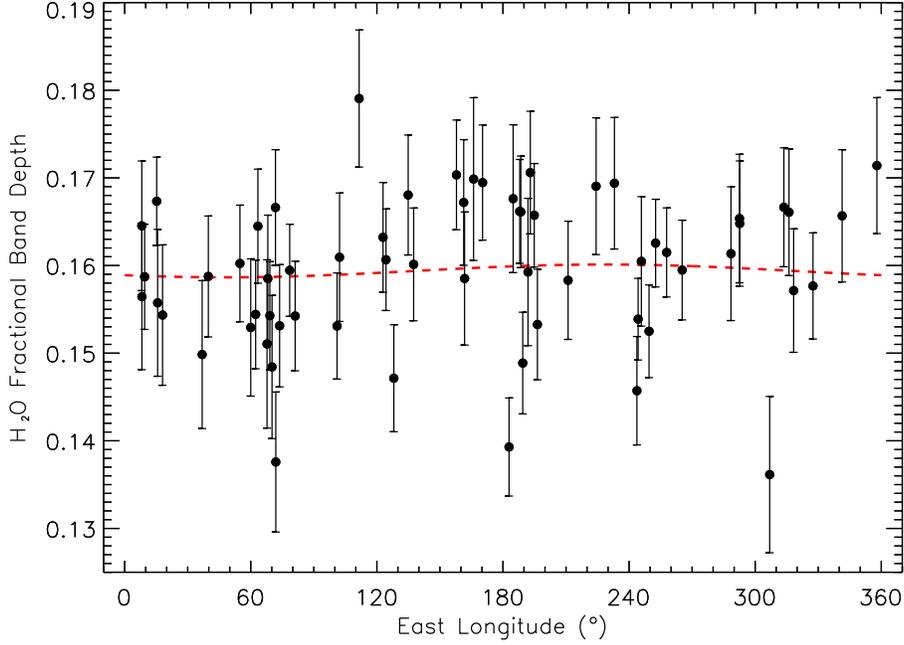}
\caption{Fractional band depth of H$_2$O ice absorption plotted against sub-observer longitude. The dashed line is a robust sinusoidal fit to the points. H$_2$O ice shows negligible variation over one Triton rotation with a large uncertainty in the longitude of peak absorption (Table 3).}
\end{center}
\end{figure}

\begin{figure}[h!]
\begin{center}
\includegraphics[scale=0.54,trim=0cm 2.5cm 1.5cm 3.25cm,clip=true]{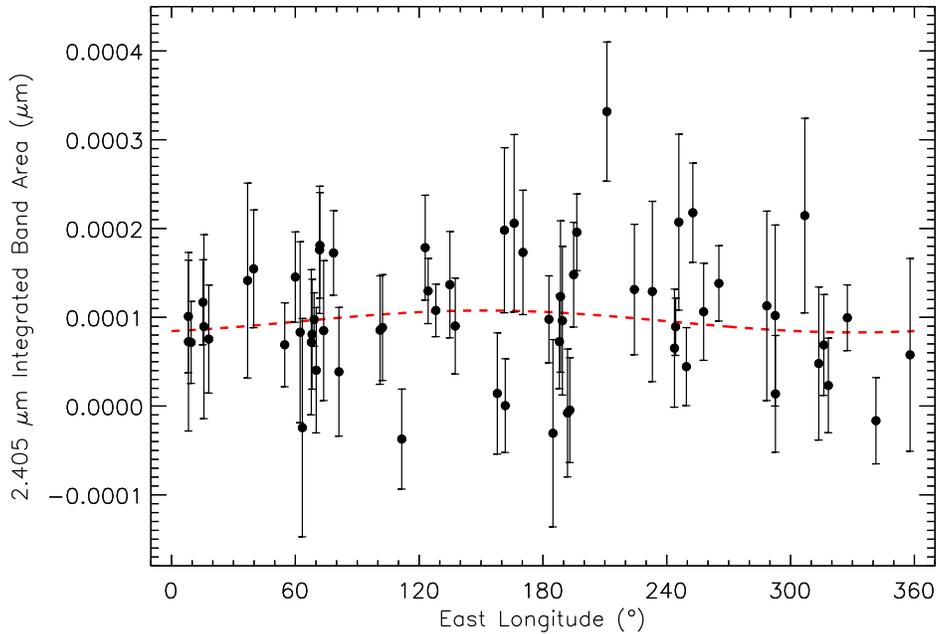}
\caption{Integrated band area of the 2.405 $\mu$m band plotted against sub-observer longitude. The dashed line is a robust sinusoidal fit to the points.  The 2.405 $\mu$m band shows negligible variation over one Triton rotation with large uncertainties in both the longitude of peak absorption and peak-to-peak amplitude (Table 3). The 2.405 $\mu$m band is consistent with ethane absorption. Negative values of integrated band area correspond to individual nightly spectra with lower signal-to-noise ratios and have no physical meaning.}
\end{center}
\end{figure}

\begin{figure}[h!]
\begin{center}
\includegraphics[scale=0.64,trim=0cm 4.5cm 1.5cm 3.5cm,clip=true]{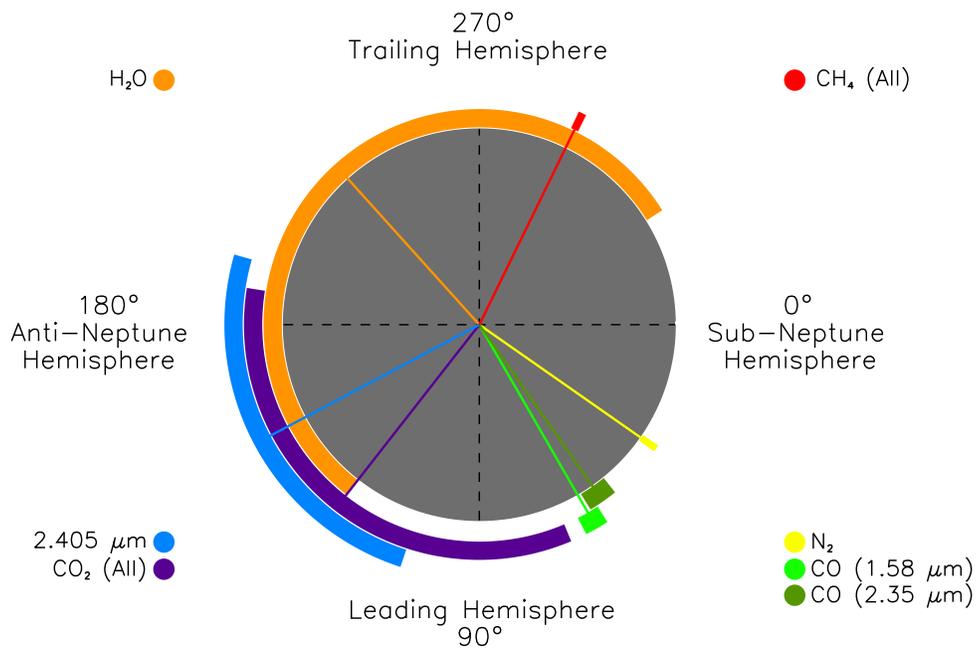}
\caption{Range of longitude of peak absorption (longitude $\pm$ uncertainty) for all ice species found on Triton's surface. The peaks from the longitudinal distributions of the combined CH$_4$ bands and the combined CO$_2$ bands are presented for those species (denoted by ``All''). Bar sizes do not represent the extent of the ices but instead show the possible location of the peak longitude for that species. Legends are in the quadrant of their respective distributions. Triton is viewed looking down on the south pole, with rotation and orbital motion counterclockwise.}
\end{center}
\end{figure}

\begin{figure}[h!]
\begin{center}
\includegraphics[scale=0.55,trim=0cm 2.5cm 1.5cm 2.5cm,clip=true]{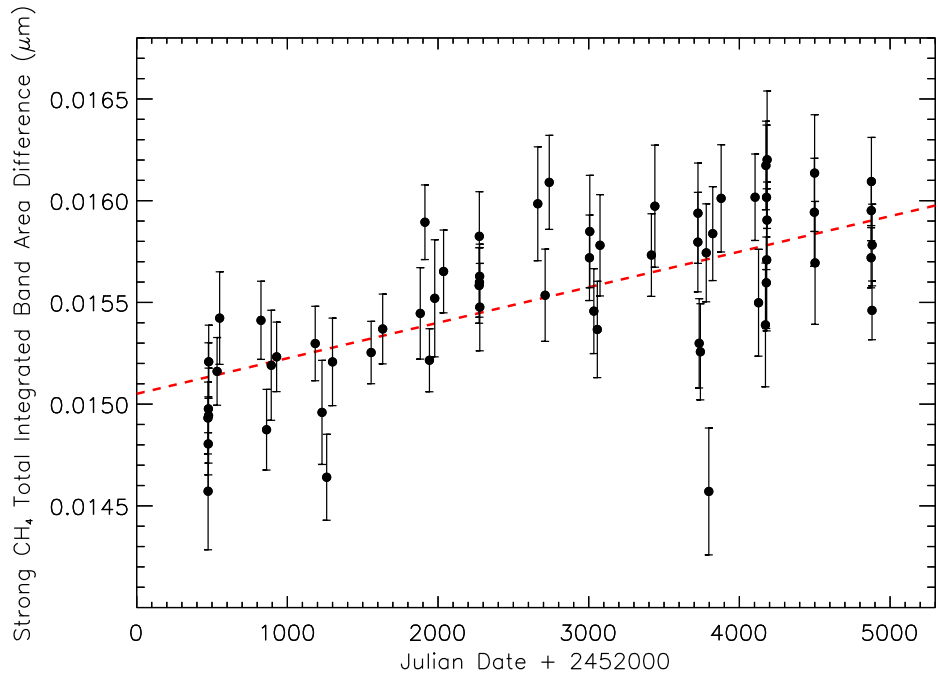}
\caption{Integrated band area of the combination of strong CH$_4$ bands plotted against Julian date. The dashed line is the robust linear fit to the points. We found an increase in absorption for the combination of strong CH$_4$ bands over the period of our observations at the 9.7-$\sigma$ level. Reduced $\chi^2$ values (Table 4) favor a sloped line over a horizontal line. The integrated band area difference was calculated by subtracting the value of the sinusoidal fit at the proper phase from the integrated band area; the mean was added to maintain the same offset as the longitudinal distribution.}
\end{center}
\end{figure}

\begin{figure}[h!]
\begin{center}
\includegraphics[scale=0.55,trim=0cm 2.5cm 1.5cm 3.25cm,clip=true]{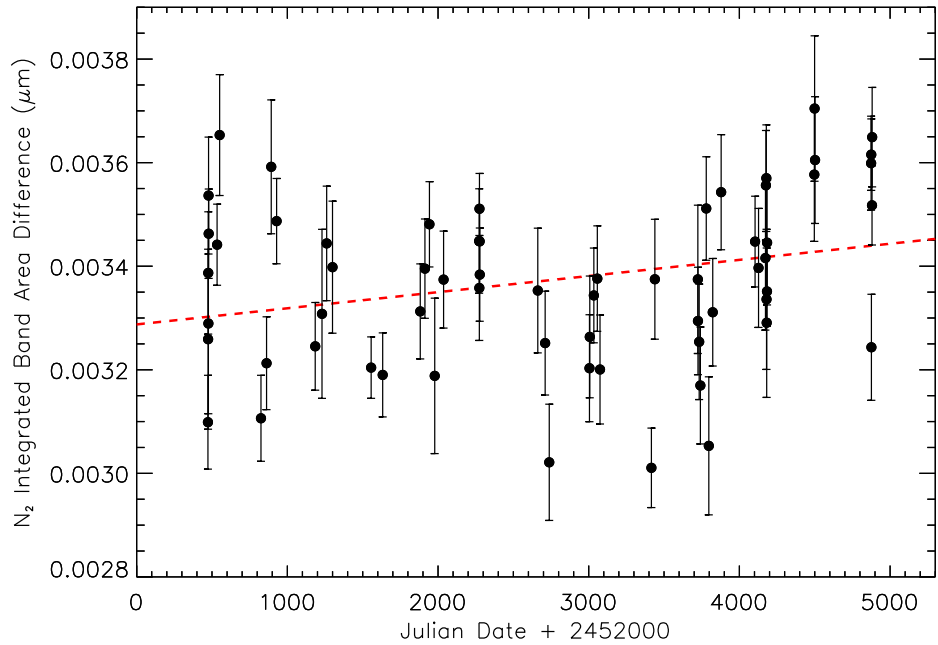}
\caption{Integrated band area of the N$_2$ band plotted against Julian date. The dashed line is the robust linear fit to the points. We found a positive slope at the 3.6-$\sigma$ level but the reduced $\chi^2$ values (Table 4) do not favor a sloped line over a horizontal line. The integrated band area difference was calculated by subtracting the value of the sinusoidal fit at the proper phase from the integrated band area; the mean was added to maintain the same offset as the longitudinal distribution.}
\end{center}
\end{figure}

\begin{figure}[h!]
\begin{center}
\includegraphics[scale=0.55,trim=0cm 2.5cm 1.5cm 3.25cm,clip=true]{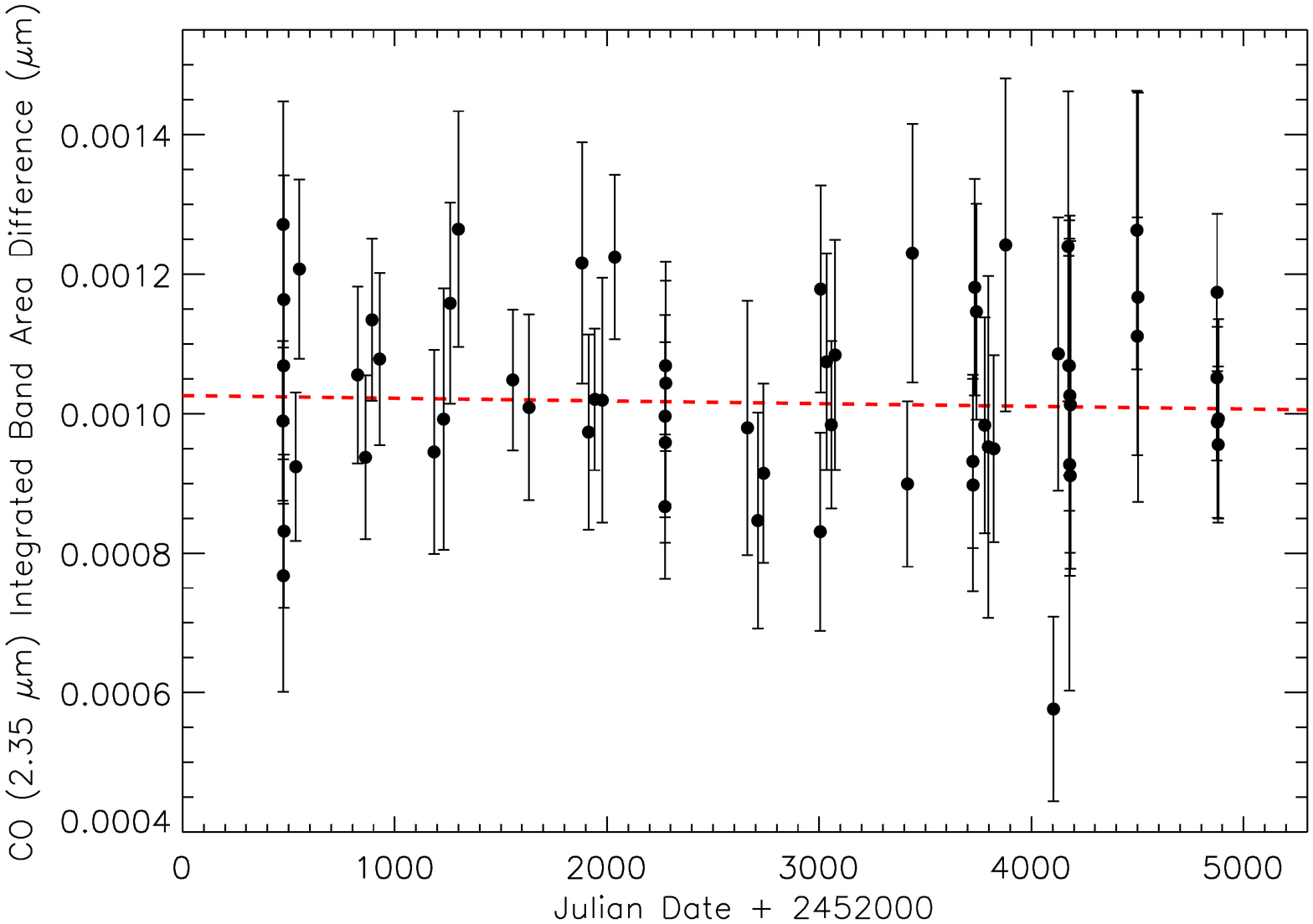}
\caption{Integrated band area of the 2.35 $\mu$m CO band plotted against Julian date. The dashed line is the robust linear fit to the points. We found a negative slope at the 0.3-$\sigma$ level and the reduced $\chi^2$ values (Table 4) do not favor a sloped line over a horizontal line. The integrated band area difference was calculated by subtracting the value of the sinusoidal fit at the proper phase from the integrated band area; the mean was added to maintain the same offset as the longitudinal distribution.}
\end{center}
\end{figure}

\begin{figure}[h!]
\begin{center}
\includegraphics[scale=0.55,trim=0cm 2.5cm 1.5cm 3.25cm,clip=true]{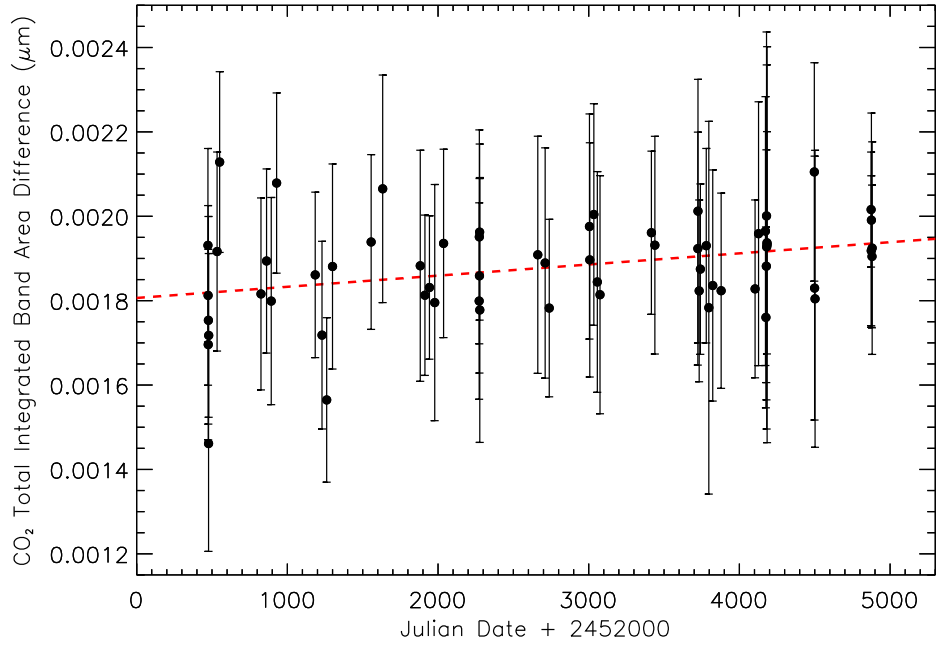}
\caption{Integrated band area of the combination of CO$_2$ bands plotted against Julian date. The dashed line is the robust linear fit to the points. We found a positive slope at the 1.3-$\sigma$ level and the reduced $\chi^2$ values (Table 4) do not favor a sloped line over a horizontal line. The integrated band area difference was calculated by subtracting the value of the sinusoidal fit at the proper phase from the integrated band area; the mean was added to maintain the same offset as the longitudinal distribution.}
\end{center}
\end{figure}

\clearpage

\section{Summary}
\onehalfspace{Between 2002 and 2014, 63 near-infrared spectra were obtained with the SpeX instrument at NASA's IRTF. Analysis of the longitudinal (zonal) and temporal distributions of various ices indicate compositional changes in the 25 years since the Voyager 2 flyby. The findings of this work are summarized below:
\begin{itemize}
\item The volatile ices, N$_2$, CO, and CH$_4$, show significant variability over one Triton rotation. N$_2$ and CO have similar longitudes of peak absorption, with CH$_4$ offset by about 90$^{\circ}$. 
\item The non-volatile ices, CO$_2$ and H$_2$O, show little variability over one Triton rotation. In fact, the longitudinal distributions for these ices are consistent with horizontal lines, suggesting a uniform distribution across the visible surface of Triton.
\item The 2.405 $\mu$m band was detected at the 4.5-$\sigma$ level in the grand average spectrum. Both non-volatile ethane (C$_2$H$_6$) and volatile $^{13}$CO absorb at 2.405 $\mu$m. The longitudinal distribution of the 2.405 $\mu$m band is 92$\pm$44$^{\circ}$ and 97$\pm$44$^{\circ}$ out of phase with the distributions for the 1.58 $\mu$m and 2.35 $\mu$m CO bands, respectively. If there is negligible fractionation of CO, this implies that the band cannot be due to $^{13}$CO. Additionally, the longitudinal distribution of the band cannot be distinguished between a sinusoid and a horizontal line, suggesting a uniform distribution across the surface, similar to the other non-volatile ices CO$_2$ and H$_2$O. This evidence is consistent with ethane absorption at 2.405 $\mu$m.
\item Over the period of the observations, absorption from the combination of strong CH$_4$ bands (1.65, 1.73, and 2.20 $\mu$m) increased. While slopes were detected above the 3.0-$\sigma$ level for N$_2$, CO (1.58 $\mu$m), and CH$_4$ (all, weak, 1.15 $\mu$m, 1.65 $\mu$m, 1.73 $\mu$m, 2.20 $\mu$m), the reduced $\chi^2$ values for these bands do not favor a sloped line over a horizontal line. The increase in CH$_4$ absorption between 2002 and 2014 may be due to volatile transport or a change in viewing geometry. Slopes for all other absorption bands were detected below the 3.0-$\sigma$ level and are therefore consistent with the null hypothesis of no temporal change.
\item The Voyager 2 flyby provided images of Triton's southern latitudes, but no spectra. The high albedo of the southern hemisphere in 1989 suggested coverage by volatile ices. The combination of volatile and non-volatile ice distributions implies a southern hemisphere currently denuded of volatile ices. Revealing the full extent of temporal changes since the Voyager 2 flyby will be the subject of future work.
\end{itemize}
\indent\indent IRTF/SpeX observations of Triton are ongoing, with five nights of spectra obtained in 2015 and a proposal submitted to continue the program in 2016. New observations are required to fully understand the processes that are altering Triton's surface. Spatially-resolved spectra of Triton would allow us to determine the latitudes of volatile and non-volatile ices. An albedo map of Triton, for comparison to previous maps made in 1989 and 2005, would be useful for quantifying the movement of ices across the surface. Information from occultations would indicate whether the surface pressure increased or decreased since the last observed occultation in 1997. Together, this information would allow us to understand Triton's surface evolution since the Voyager 2 flyby in 1989 and to predict future changes.}

\section{Acknowledgements}
\onehalfspace{We graciously thank the staff of the IRTF for their assistance over the past 12+ years, especially, W. Golisch, D. Griep, P. Sears, E. Volquardsen, J. T. Rayner, A. T. Tokunaga, and B. Cabreira. A special thanks to the reviewers whose comments significantly improved this paper. We wish to recognize and acknowledge the significant cultural role and reverance of the summit of Mauna Kea within the indigenous Hawaiian community and to express our appreciation for the opportunity to observe from this special mountain. This work was funded by NASA PAST NNX13AG06G and NASA NESSF 14-PLANET14F-0045.}

\section*{Appendix}
\onehalfspace{We derive here an equation for the calculation of the projected solid angle of a portion of a sphere. An infinitesimal solid angle (d$\Omega$) on the surface of a sphere can be calculated from the product of the infinitesimal latitude (d$\theta$) and longitude (d$\phi$), and the cosine of the latitude:
\begin{equation}
\mathrm{d}\Omega=\mathrm{d}\theta \mathrm{d}\phi \mathrm{cos}\theta. \tag{A1}
\end{equation}
The cos$\theta$ term is necessary since the solid angle of the poles themselves is zero. Integrating Eq. (A1) over the entire surface of the sphere ($\phi$ from 0 to 2$\pi$ and $\theta$ from $-\frac{\pi}{2}$ to $\frac{\pi}{2}$) results in a solid angle of 4$\pi$, as expected.\\
\indent For an observer, the emission angle ($\chi$) from each d$\Omega$ affects the calculation of the surface area visible at any given time. The projected solid angle ($\Omega_{proj}$) is the integral of the product of the solid angle and the cosine of the emission angle:
\begin{equation}
\Omega_{proj}=\int_{\phi_1}^{\phi_2} \int_{\theta_1}^{\theta_2} \mathrm{cos}\chi \mathrm{cos}\theta \mathrm{d}\theta \mathrm{d}\phi. \tag{A2}
\end{equation}
The total projected solid angle has a value of $\pi$ and can be visualized as an arbitrary hemisphere projected onto a circular disk of the same radius.\\
\indent Now consider two vectors, $\widehat{r}$ and $\widehat{r_0}$, that represent the direction from the center of the sphere to a surface element d$\Omega$ and the direction from the center of the sphere to the observer, respectively. The $x$, $y$, and $z$ components of $\widehat{r}$ and $\widehat{r_0}$ are measured from the center of the sphere as seen by the observer. In vector notation, we have:
\begin{equation}
\widehat{r}=(\mathrm{cos}\theta \mathrm{cos}\phi, \mathrm{cos}\theta \mathrm{sin}\phi, \mathrm{sin}\theta) \tag{A3}
\end{equation}
and
\begin{equation}
\widehat{r_0}=(\mathrm{cos}\theta_0 \mathrm{cos}\phi_0, \mathrm{cos}\theta_0 \mathrm{sin}\phi_0, \mathrm{sin}\theta_0). \tag{A4}
\end{equation}
The emission angle is defined by the dot product of Eq. (A3) and (A4):
\begin{equation}
\widehat{r}\cdot\widehat{r_0}=|\widehat{r}||\widehat{r_0}|\mathrm{cos}\chi \Rightarrow \mathrm{cos}\chi=\frac{\widehat{r}\cdot\widehat{r_0}}{|\widehat{r}||\widehat{r_0}|} \tag{A5} \nonumber
\end{equation}
Since $\widehat{r}$ and $\widehat{r_0}$ are unit vectors, their moduli are 1. Therefore, cos$\chi$ is equal to the dot product of the two unit vectors:
\begin{equation}
\mathrm{cos}\chi=\widehat{r}\cdot\widehat{r_0}=\mathrm{cos}\theta\mathrm{cos}\phi\mathrm{cos}\theta_0\mathrm{cos}\phi_0+\mathrm{cos}\theta\mathrm{sin}\phi\mathrm{cos}\theta_0\mathrm{sin}\phi_0+\mathrm{sin}\theta\mathrm{sin}\theta_0 \tag{A5} \nonumber
\end{equation}
The sub-observer longitude is chosen to be zero by rotational symmetry, simplifying the expression for cos$\chi$:
\begin{equation}
\mathrm{cos}\chi=\mathrm{cos}\theta\mathrm{cos}\phi\mathrm{cos}\theta_0+\mathrm{sin}\theta\mathrm{sin}\theta_0 \tag{A6}
\end{equation}
\indent Plugging Eq. (A6) into Eq. (A2) yields the projected area in terms of latitude, longitude, sub-observer latitude, and sub-observer longitude:
\begin{equation}
\Omega_{proj}=\int_{\phi_1}^{\phi_2} \int_{\theta_1}^{\theta_2} \mathrm{cos}\theta[\mathrm{cos}\theta\mathrm{cos}\phi\mathrm{cos}\theta_0+\mathrm{sin}\theta\mathrm{sin}\theta_0] \mathrm{d}\theta \mathrm{d}\phi. \tag{A7}
\end{equation}
\indent In this work we implemented a numerical calculation of Eq. (A7) in IDL. We started by considered all longitudes ($\phi_1$=0 and $\phi_2$=2$\pi$) for a chosen range of latitudes, then ignored any infinitesimal surface areas with a corresponding value of cos$\chi$ that was negative. By dividing the resulting projected area by $\pi$, the value of the total projected area, we obtained a percentage of the projected area due to the region in question.}

\section*{References}
\noindent Agnor, C.B., Hamilton, D.P., 2006. Neptune's capture of its moon Triton in a binary-planet gravitational encounter. Nature 441, 192-194.\\
\\
Barth, E.L., Toon, O.B., 2003. Microphysical modeling of ethane ice clouds in Titan's atmosphere. Icarus 162, 94-113.\\
\\
Bauer, J.M., Buratti, B.J., Li, J.-Y., Mosher, J.A., Hicks, M.D., Schmidt, B.E., Goguen, J.D., 2010. Direct detection of seasonal changes on Triton with Hubble Space Telescope. ApJL 723, L49-L52.\\
\\
Broadfoot, A.L., et al., 1989. Ultraviolet spectrometer observations of Neptune and Triton. Science 246, 1459-1466.\\
\\
Brown, R.H., Cruikshank, D.P., Veverka, J., Helfenstein, P., Eluszkiewicz, J., 1995. Surface composition and photometric properties of Triton. In: Cruikshank, D.P. (Ed.), Neptune and Triton. University of Arizona Press, Tucson, pp. 991-1030.\\
\\
Buie, M.W., Bus, S.J., 1992. Physical observations of (5145) Pholus. Icarus 100, 288-294.\\
\\
Buratti, B.J., Goguen, J.D., Gibson, J., Mosher, J., 1994. Historical photometric evidence for volatile migration on Triton. Icarus 110, 303-314.\\
\\
Chen, F.Z., Wu, C.Y.R., 2004. Temperature-dependent photoabsorption cross sections in the VUV-UV region. I. Methane and ethane. J. Quant. Spectrosc. Radiat. Trans. 85, 195-209.\\
\\
Cruikshank, D.P., Roush, T.L., Owen, T.C., Geballe, T.R., de Bergh, C., Schmitt, B., Brown, R.H., Bartholomew, M.J., 1993. Ices on the surface of Triton. Science 261, 742-745.\\
\\
Cruikshank, D.P., Schmitt, B., Roush, T.L., Owen, T.C., Quirico, E., Geballe, T.R., de Bergh, C., Bartholomew, M.J., Dalle Ore, C.M., Dout{\'e}, S., Meier, R., 2000. Water ice on Triton. Icarus 147, 309-316.\\
\\
Cruikshank, D.P., Mason, R.E., Dalle Ore, C.M., Bernstein, M.P., Quirico, E., Mastrapa, R.M., Emery, J.P., Owen, T.C., 2006. Ethane on Pluto and Triton. Bull. Am.  Astron. Soc. 38, 518.\\
\\
DeMeo, F.E., Dumas, C., de Bergh, C., Protopapa, S., Cruikshank, D.P., Geballe, T.R., Alvarez-Candal, A., Merlin, F., Barucci, M.A., 2010. A search for ethane on Pluto and Triton. Icarus 208, 412-424.\\
\\
Elliot, J.L., et al., 1998. Global warming on Triton. Nature 393, 765-767.\\
\\
Fray, N., Schmitt, B., 2009. Sublimation of ices of astrophysical interest: A bibliographic review. P\&SS 57, 2053-2080.\\
\\
Grundy, W.M., Young, L.A., Stansberry, J.A., Buie, M.W., Olkin, C.B., Young, E.F., 2010. Near-infrared spectral monitoring of Triton with IRTF/SpeX II: Spatial distribution and evolution of ices. Icarus 205, 594-604.\\
\\
Grundy, W.M., Olkin, C.B., Young, L.A., Buie, M.W., Young, E.F., 2013. Near-infrared spectral monitoring of Pluto's ices: Spatial distribution and secular evolution. Icarus 223, 710-721.\\
\\
Gurrola, E.M., 1995. Interpretation of Radar Data from the Icy Galilean Satellites and Triton. Ph.D. thesis, Stanford University.\\
\\
Hansen, C.J., Paige, D.A., 1992. A thermal model for the seasonal nitrogen cycle on Triton. Icarus 99, 273-288.\\
\\
Herbert, F., Sandel, B.R., 1991. CH$_4$ and haze in Triton's lower atmosphere. JGR Supp. 96, 19241-19252.\\
\\
Hicks, M.D., Buratti, B.J., 2004. The spectral variability of Triton from 1997-2000. Icarus 171, 210-218.\\
\\
Holler, B.J., Young, L.A., Grundy, W.M., Olkin, C.B., Cook, J.C., 2014. Evidence for longitudinal variability of ethane ice on the surface of Pluto. Icarus 243, 104-110.\\
\\
Horne, K., 1986. An optimal extraction algorithm for CCD spectroscopy. PASP 98, 609-617.\\
\\
Houk, N., Smith-Moore, M., 1988. Michigan Catalogue of Two-dimensional Spectral Types for the HD Stars. Volume 4, Declinations -26$^{\circ}$.0 to -12$^{\circ}$.0. Department of Astronomy, University of Michigan, Ann Arbor, MI.\\
\\
Houk, N., Swift, C., 1999. Michigan Catalogue of Two-dimensional Spectral Types for the HD Stars; vol. 5. Department of Astronomy, University of Michigan, Ann Arbor, MI.\\
\\
Ingersoll, A.P., 1990. Dynamics of Triton's atmosphere. Nature 344, 315-317.\\
\\
Johnson, R.E., Oza, A., Young, L.A., Volkov, A.N., Schmidt, C., 2015. Volatile loss and classification of Kuiper Belt Objects. ApJ 809, 43-51.\\
\\
Krasnopolsky, V.A., Cruikshank, D.P., 1995. Photochemistry of Triton's atmsophere and ionosphere. JGR 100, 21271-21286.\\
\\
Lara, L.M., Ip, W.-H., Rodrigo, R., 1997. Photochemical Models of Pluto's Atmosphere. Icarus 130, 16-35.\\
\\
Lellouch, E., de Bergh, C., Sicardy, B., Ferron, S., K{\"a}ufl, H.-U., 2010. Detection of CO in Triton's atmosphere and the nature of surface-atmosphere interactions. A\&A 512, L8-L13.\\
\\
McKinnon, W.B., Lunine, J.I., Banfield, D., 1995. Origin and evolution of Triton. In: Cruikshank, D.P. (Ed.), Neptune and Triton. University of Arizona Press, Tucson, pp. 807-877.\\
\\
Moore, J.M., Spencer, J.R., 1990. Koyaanismuuyaw - The hypothesis of a perenially dichotomous Triton. GRL 17, 1757-1760.\\
\\
Moore, M.H., Hudson, R.L., 2003. Infrared study of ion-irradiated N$_2$-dominated ices relevant to Triton and Pluto: formation of HCN and HNC. Icarus 161, 486-500.\\
\\
Neckel, H., 1986. The ``bright stars'' with UBV-colors close to those of the Sun. A\&A 169, 194-200.\\
\\
Pourbaix, D., Tokovinin, A.A., Batten, A.H., Fekel, F.C., Hartkopf, W.I., Levato, H., Morrell, N.I., Torres, G., Udry, S., 2004. S$^9_{\mathrm{B}}$: The ninth catalogue of spectroscopic binary orbits. A\&A 424, 727-732.\\
\\
Prokhvatilov, A.I., Yantsevich, L.D., 1983. X-ray investigations of the equilibrium phase diagram of CH$_4$-N$_2$ solid mixtures. Sov. J. Low Temp. Phys. 9, 94-97.\\
\\
Protopapa, S., Grundy, W.M., Tegler, S.C., Bergonio, J.M., 2015. Absorption coefficients of the methane-nitrogen binary ice system: Implications for Pluto. Icarus 253, 179-188.\\
\\
Rages, K., Pollack, J.B., 1992. Voyager imaging of Triton's clouds and hazes. Icarus 99, 289-301.\\
\\
Rayner, J.T., Toomey, D.W., Onaka, P.M., Denault, A.J., Stahlberger, W.E., Watanabe, D.Y., Wang, S.I., 1998. SpeX: A medium-resolution IR spectrograph for IRTF. Proc. SPIE 3354, 468-479.\\
\\
Rayner, J.T., Toomey, D.W., Onaka, P.M., Denault, A.J., Stahlberger, W.E., Vacca, W. D., Cushing, M.C., Wang, S., 2003. SpeX: A medium-resolution 0.8-5.5 micron spectrograph and imager for the NASA Infrared Telescope Facility. Publ. Astron. Soc. Pacific 115, 362-382.\\
\\
Schaller, E.L., Brown, M.E., 2007. Volatile loss and retention on Kuiper Belt Objects. ApJ 659, L61-L63.\\
\\
Scott, P.C., Asplund, M., Grevesse, N., Sauval, A.J., 2006. Line formation in solar granulation. VII. CO lines and the solar C and O isotopic abundances. A\&A 456, 675-688.\\
\\
Sicardy, B., Boissel, Y., Colas, F., Doressoundiram, A., Lecacheux, J., Widemann, T., Frappa, E., Bath, K.-L., Beisker, W., Bernard, O., 2008. The Triton stellar occultation of 21 May 2008. EPSC, M{\"u}nster, Germany, 21-25 September 2008.\\
\\
Smith, B.A., et al., 1989. Voyager 2 at Neptune: Imaging science results. Science 246, 1422-1449.\\
\\
Stone, E.C., Miner, E.D., 1989. The Voyager 2 encounter with the Neptunian system. Science 246, 1417-1421.\\
\\
Tegler, S.C., Grundy, W.M., Olkin, C.B., Young, L.A., Romanishin, W., Cornelison, D.M., Khodadadkouchaki, R., 2012. Ice mineralogy across and into the surfaces of Pluto, Triton, and Eris. ApJ 751, 76.1-10.\\
\\
Trafton, L., 1984. Large seasonal variations in Triton's atmosphere. Icarus 58, 312-324.\\
\\
Tryka, K.A., Brown, R.H., Anicich, V., Cruikshank, D.P., Owen, T.C., 1993. Spectroscopic determination of the phase composition and temperature of nitrogen ice on Triton. Science 261, 751-754.\\
\\
Tyler, G.L., Sweetnam, D.N., Anderson, J.D., Borutzki, S.E., Campbell, J.K., Kursinski, E.R., Levy, G.S., Lindal, G.F., Lyons, J.R., Wood, G.E., 1989. Voyager radio science observations of Neptune and Triton. Science 246, 1466-1473.\\
\\
Vetter, M., Jodl, H.-J., Brodyanski, A., 2007. From optical spectra to phase diagrams--the binary mixture N$_2$-CO. Low. Temp. Phys. 33, 1052-1060.\\

\end{document}